\documentclass[12pt]{article}
\usepackage{amssymb,amsmath,amscd}
\usepackage{epsfig}
\usepackage{epstopdf}
\usepackage{latexsym}
\usepackage{graphicx}
\usepackage{booktabs}
\usepackage{bbm}
\usepackage[margin=15pt,small]{caption}
\usepackage{subcaption}
\usepackage{enumitem}
\usepackage{float}
\usepackage[toc]{appendix}
\usepackage[numbers,compress]{natbib}
\usepackage{tikz}
\usetikzlibrary{positioning}
\usepackage{array,longtable}
\usepackage{autobreak}
\usepackage{listings}
\allowdisplaybreaks

\definecolor{cardinal}{rgb}{0.6,0,0}
\definecolor{darkgreen}{rgb}{0.1,0.4,0}
\definecolor{golden}{rgb}{0.92, 0.7, 0}
\definecolor{midnight}{rgb}{0, 0, 0.5}
\definecolor{darkblue}{rgb}{0.3,0.3,0.7}
\definecolor{darkred}{rgb}{0.7,0,0}

\usepackage[T1]{fontenc}
\usepackage{lmodern}

\usepackage{color}
\usepackage{datetime}

\usepackage{hyperref}
\hypersetup{
  colorlinks=true,
  linkcolor=darkblue,
  citecolor=darkred,
  urlcolor=darkgreen,
  linktocpage=true,
  linktoc=page
}

\setlist{nolistsep}

\newcolumntype{L}{>{$}l<{$}}
\allowdisplaybreaks

\ifpdf
\fi

\let\oldbibliography\thebibliography
\renewcommand{\thebibliography}[1]{\oldbibliography{#1}
\setlength{\itemsep}{0pt}} 

\newcommand*{\boxedcolor}{red}
\makeatletter
\renewcommand{\boxed}[1]{\textcolor{\boxedcolor}{%
  \fbox{\normalcolor\m@th$\displaystyle#1$}}}
\makeatother

\def\coeff#1#2{\relax{\textstyle {#1 \over #2}}\displaystyle}

\def\ZZ{\mathbb{Z}}

\def\eql{~=~}

\def\cals#1{\mathcal{#1}}

\def\eql{=}

\def\Tr{{\rm Tr}\,}

\def\RR{\mathbb{R}}

\def\GL{{\rm GL}}
\def\SL{{\rm SL}}

\def\SO{{\rm SO}}

\def\SU{{\rm SU}}

\def\sl{\frak{sl}}

\def\so{\frak{so}}
\def\su{\frak{su}}

\def\gl{\frak{gl}}

\def\bfs#1{{\boldsymbol{#1}}}

\def\cals#1{\mathcal{#1}}

\def\T#1#2#3#4#5#6#7{{\tt T#1#2#3#4#5#6#7}}

\def\fracc#1#2{#1/#2}

\numberwithin{equation}{section} 
\numberwithin{table}{section}
\numberwithin{figure}{section}

\newcommand{\e}{\mathrm{e}}

\newcommand{\be}{\begin{equation}}
\newcommand{\ee}{\end{equation}}
\newcommand{\bea}{\begin{eqnarray}}
\newcommand{\eea}{\end{eqnarray}}

\newcommand{\f}[2]{\frac{#1}{#2}}

\newcommand{\U}{\text{U}}

\begin{document}  

\begin{titlepage}
 
\medskip
\begin{center} 
{\Large \bf A  Cornucopia of AdS$_5$ Vacua}

\bigskip
\bigskip
\bigskip
\bigskip

{\bf Nikolay Bobev,${}^{\rm A}$ Thomas Fischbacher,${}^{\rm d}$ \\Fri\dh rik Freyr Gautason,${}^{\rm A,S}$ and Krzysztof Pilch${}^{5}$   \\ }
\bigskip
${}^{\rm A}$
Instituut voor Theoretische Fysica, KU Leuven,\\ 
Celestijnenlaan 200D, B-3001 Leuven, Belgium
\vskip 5mm
${}^{\rm d}$ Google Research\\
Brandschenkestrasse 110, 8002 Z\"urich, Switzerland
\vskip 5mm
${}^{\rm S}$ University of Iceland, Science Institute\\
Dunhaga 3, 107 Reykjav\'ik, Iceland
\vskip 5mm
${}^{5}$ Department of Physics and Astronomy \\
University of Southern California \\
Los Angeles, CA 90089, USA  \\
\bigskip
\tt{nikolay.bobev@kuleuven.be, ffg@kuleuven.be, tfish@google.com, pilch@usc.edu}  \\
\end{center}

\bigskip
\bigskip

\begin{abstract}

{\noindent  We report on a systematic search for AdS$_5$ vacua corresponding to critical points of the potential in the five-dimensional $\mathcal{N}=8$ $\SO(6)$ gauged supergravity. By employing  Google's TensorFlow Machine Learning library, we find the total of 32 critical points including 5~previously known ones. All 27  new critical points are non-supersymmetric. We compute the  mass spectra of scalar fluctuatons for all points and find that the  non-supersymmetric AdS$_5$ vacua are perturbatively unstable. Many of the new critical points can be found analytically within consistent truncations of the $\cals N=8$ supergravity with respect to discrete subgroups of the $\rm S(O(6)\times \GL(2,\RR))$ symmetry of the potential. In particular, we discuss in detail a $\ZZ_2^3$-invariant truncation with 10 scalar fields and 15 critical points. We also compute explicitly the scalar potential in a $\ZZ_2^2$-invariant extension of that truncation to 18~scalar fields  and reproduce  17 of the 32 critical points from the numerical search. Finally, we show that the full potential as a function of  42 scalar fields can be studied analytically   using the so-called solvable parametrization. In particular, we find that all critical points lie in a $\ZZ_2$-invariant subspace spanned by 22 scalar fields.
}
\end{abstract}

\end{titlepage}

\newpage
\setcounter{tocdepth}{2}
\tableofcontents

\section{Introduction}
\label{sec:Introduction}

The AdS/CFT correspondence is deeply rooted in string theory and its low-energy supergravity limits. Therefore, it is important to understand fully  the landscape of consistent AdS backgrounds in string theory. A fruitful strategy has been to identify a consistent Kaluza-Klein (KK) truncation of a ten- or eleven-dimensional supergravity  to lower $d$ dimensions and to study critical points of the  scalar potential in the resulting  gauged supergravity. Each critical point  with  a negative value  of the potential leads  to an AdS$_d$ solution and  thus  candidate AdS background  of string theory.

Our goal in this paper is to use a mixture of  old  analytic and modern numerical methods to search systematically for critical points of the scalar potential in $\cals N=8$ SO(6) gauged supergravity in five dimensions \cite{Gunaydin:1984qu,Gunaydin:1985cu,Pernici:1985ju}. This is interesting for several reasons. First, there is now a complete, constructive  proof that this five-dimensional supergravity is a consistent KK truncation of type IIB string theory on $S^5$ \cite{Khavaev:1998fb,Cvetic:2000nc,Pilch:2000ue,Lee:2014mla,Baguet:2015sma}. In particular, this means that all  AdS$_5$ vacua corresponding to critical points of the supergravity potential can be uplifted to  AdS solutions of string theory. 
Secondly, the problem should be amenable to   similar computational techniques  based on Machine Learning that were successfully applied in \cite{Comsa:2019rcz} to find hundreds of new critical points of the scalar potential in the de Wit-Nicolai SO(8) gauged supergravity in four dimensions \cite
{deWit:1982bul}. Finally, by extrapolating the results in \cite{Bobev:2019dik}, it is natural to expect that  a large fraction of the critical points might be accessible analytically, or semi-analytically,  within a suitable truncation  with respect to a discrete subgroup of the full symmetry group of the theory.

Through holography, the  $\SO(6)$ gauged supergravity has been an indispensable tool for studying the $\mathcal{N}=4$ SYM theory and its deformations. Indeed,   AdS$_5$ vacua are  dual to conformal fixed points obtained by deforming   $\mathcal{N}=4$ SYM  and domain wall solutions between these critical points are dual to RG flows between the CFTs \cite{Girardello:1998pd,Distler:1998gb,Freedman:1999gp}. From this perspective, one would also like  to determine the stability of those vacua. If an AdS$_5$ solution is supersymmetric, it is necessarily stable \cite{Gibbons:1983aq} and the dual CFT is unitary. However, if there are scalar fluctuations with negative masses violating the Breitenlohner-Freedman (BF) bound  \cite{Breitenlohner:1982jf}, the  dual operators have complex dimensions and the dual CFT is not unitary.
In fact, it has been argued in \cite{Ooguri:2016pdq} using the Weak Gravity Conjecture \cite{ArkaniHamed:2006dz} that all non-supersymmetric vacua in string theory should be unstable. The violation of the BF bound for a given AdS solution is then the simplest sign of that instability.

It is perhaps surprising that  not much progress has been made in classifying  AdS$_5$ vacua  of the $\SO(6)$ gauged supergravity since the initial discovery in 1998 of five critical points listed in Table~\ref{tbl:KPW}\footnote{Following the convention for labelling critical points in four-dimensional supergravity \cite{Fischbacher:2011jx}, we propose to denote the points in five dimensions according to the value of the first seven digits in their cosmological constant  by $\tt Tn_1n_2n_3n_4n_5n_6n_7$.}  in an $\SU(2)$-invariant sector of the theory \cite{Khavaev:1998fb}. One reason might be that the Leigh-Strassler  analysis \cite{Leigh:1995ep} of $\cals N=1$  deformations of $\cals N=4$ SYM   suggests that there should be no other supersymmetric critical points beyond the $\cals N=8$ point, $\tt T0750000$, and the $\cals N=2$ point, $\tt T0839947$, already found in \cite{Khavaev:1998fb}. The other three points in Table~\ref{tbl:KPW} are non-supersymmetric and  perturbatively unstable as discussed further in Appendix~\ref{appendixC}. It is then reasonable to expect that any missing point is non-supersymmetric and  thus perturbatively unstable as well. Note, however, that the latter need not be necessarily true given that there is a perturbatively stable yet non-supersymmetric $\rm SO(3)\times SO(3)$-invariant AdS$_4$ solution in four dimensions \cite{Warner:1983du,Fischbacher:2010ec},\footnote{However, it has been shown recently that this solution is unstable in string theory due to brane-jet instability \cite{Bena:2020misc} and higher KK-modes violating the BF bound \cite{Malek:2020misc}.}   and there are multiple examples of perturbatively stable AdS$_3$ vacua in three-dimensional supergravities~\hbox{\cite{Fischbacher:2002fx,Fischbacher:2008zu}}.

\begin{table}[t]
\renewcommand{\arraystretch}{1.0}
\begin{center}
\scalebox{1}{
\begin{tabular}{@{\extracolsep{10 pt}} c c  c  c   c   }
\toprule
\noalign{\smallskip}
Point &  Symmetry  &    $\cals P_*$  & SUSY & BF Stability  \\
\noalign{\smallskip}
\midrule
\noalign{\medskip}
$\tt T0750000$ & SO(6)  &   $-{3\over 4} $   & $\cals N=8$ & S   \\[6 pt]
$\tt T0780031$ & $\rm SO(5)$     &   $-{3^{5/3}\over 8}$ & -- & U    \\[6 pt]
$\tt T0839947$ &   $\rm SU(2)\times U(1)$   &    $-{2^{4/3}\over 3}$ & $\cals N=2$ & S  \\[6 pt]
$\tt T0843750$ &   $\rm SU(3)$  &    $-{27\over 32}$ & -- & U  \\[6 pt]
$\tt T0870297$ & $\rm SU(2)\times U(1)^2$  & $-{3\over 8}\left({25\over 2}\right)^{1/3}$ & -- & U   \\[6 pt]
\noalign{\smallskip}
\bottomrule
\end{tabular}
}
\caption{\label{tbl:KPW} The $\rm SU(2)$-invariant extrema \cite{Khavaev:1998fb}.}
\end{center}
\end{table}

Given the large number of known  critical points of the scalar potentials in maximal gauged supergravities in three \cite{Fischbacher:2002fx,Fischbacher:2008zu} and four dimensions \cite{Warner:1983du,Warner:1983vz,Fischbacher:2009cj,Fischbacher:2010ec,Fischbacher:2011jx,Borghese:2013dja,Comsa:2019rcz,Bobev:2019dik}, it is to be expected that there are comparably many AdS$_5$ vacua of the five-dimensional $\SO(6)$ gauged supergravity beyond  the  ones in Table~\ref{tbl:KPW}. It is the lower symmetry (less than $\SU(2)$) of these vacua that makes looking for them a challenging problem. 

Recall that the  potential of the $\cals N=8$ $d=5$ supergravity is a function on the 42-dimensional scalar manifold, which is a coset of the maximally noncompact group $\rm E_{6(6)}$ modded by its compact subgroup, $\rm USp(8)$. In the conventions of \cite{Gunaydin:1985cu},\footnote{See also Appendix~\ref{appConv}.} 
the potential  can be written as
\begin{equation}\label{Pott}
\cals P\eql -{1\over 32}\,g^2\,\left[ \,2(W_{ab})^2-(W_{abcd})^2\,\right]\,,
\end{equation}
which  looks deceptively simple until fully unpacked. Indeed, the  $W_{ab}$ and $W_{abcd}$ tensors  are quadratic in the components of the scalar 27-bein, $\cals V=(\cals V^{IJab},\cals V_{I\alpha}{}^{ab})$, which, modulo a linear transformation, is a group element of $\rm E_{6(6)}$ obtained by exponentiating  non-compact elements, $\Phi=\sum\phi_A T_A$, of the Lie algebra  $\frak e_{6(6)}$, where   $T_A$ are some fixed generators and $\phi_A$  are the 42 scalars fields. It follows from the construction of the $W$-tensors that the potential is manifestly invariant  under the $\rm SO(6)$ gauge symmetry acting on the  $I,J=1,\ldots,6$ indices as well as the axion-dilaton $\SL(2,\RR)$ that acts on the $\alpha=7,8$ index of the 27-bein. This reduces the number of independent degrees of freedom in \eqref{Pott} to $42-15-3=24$. In fact, by including discrete symmetries one can show that the  actual symmetry is $\rm S(O(6)\times GL(2,\RR))$ \cite{Pilch:2000fu,Bobev:2016nua}, which we will exploit in Section~\ref{sec:truncation}.  When viewed as a function on $\rm E_{6(6)}$, the potential \eqref{Pott} is also invariant 
under local  $\rm USp(8)$ transformations  acting on the $a,b=1,\ldots,8$ indices of the 27-bein, but that symmetry is already fixed by the $\rm USp(8)$ gauge choice in $\Phi$.

The problem now is to compute the potential, $\cals P(\phi_A)$, as an explicit  function of the scalar fields and then find its critical points. Analytically, this cannot be done in full generality. A time-tested method, first used by Warner \cite{Warner:1983du,Warner:1983vz} in four dimensions, is to truncate the potential of interest to a smaller number of fields that are invariant under some subgroup, $G$, of the full symmetry group of the theory. The critical points of the truncated potential are then automatically critical points of the full potential. For a judicious choice of the subgroup,  $G$, one may end up with an analytically tractable problem leading to a potential with  new critical points. As we  discuss briefly in Section~\ref{sec:truncation},  this method has not been too successful thus far in five dimensions beyond the original analysis in \cite{Khavaev:1998fb}. The scalar potentials in various truncations considered  over the years in the literature either did not include new critical points or were deemed too complicated to attempt extremization.

Another  way to make progress is to attack the problem numerically. This has been initiated about ten years ago by one of the authors and resulted in around 40 new  AdS$_4$ vacua \cite{Fischbacher:2008zu,Fischbacher:2009cj,Fischbacher:2010ki,Fischbacher:2010ec,Fischbacher:2011jx}  in the de Wit-Nicolai  $\SO(8)$ gauged supergravity for the total of 50 critical points known in 2013.\footnote{Those include the 7 original points found in the ``classic period'' \cite{Warner:1983du,Warner:1983vz} and one futher point    in \cite{Borghese:2013dja}.}  
Recently, a more powerful numerical code using Machine Learning (ML) and Google's TensorFlow libraries \cite{Abadi2016} was developed in \cite{Comsa:2019rcz} and led to the total of 194 points that include  2 additional ones found in the follow up analytic work  \cite{Bobev:2019dik}. It is rather straightforward to port the  ML code included with  \cite{Comsa:2019rcz}  from four to five dimensions and, in fact, considerably simplify it using the new publicly available  TensorFlow2 libraries\footnote{Cf. \href{https://blog.tensorflow.org/2019/09/tensorflow-20-is-now-available.html}{\texttt{https://blog.tensorflow.org/2019/09/tensorflow-20-is-now-available.html}}} as well as by exploiting symmetries of the potential. 

By performing a systematic, numerical search using the new ML code, we find the total of {32 AdS$_5$} vacua in $\cals N=8$ $d=5$ SO(6) gauged supergravity. Those include the 5   classic  ones in Table~\ref{tbl:KPW}. We also compute the gravitini  and  scalar spectra at each point, which are needed to determine unbroken supersymmetries and the BF (in)stability. We find that all {27~new points} are non-supersymmetric, which is compatible with the expectation that the dual $\mathcal{N}=4$ SYM theory does not admit  relevant deformations, apart from the one in \cite{Leigh:1995ep}, which lead to interacting supersymmetric CFTs. All new points  have BF unstable scalar modes, which is perhaps disappointing, but not unexpected. Hence our results  further support  the instability conjecture for non-supersymmetric AdS vacua in string theory \cite{Ooguri:2016pdq}.

It turns out that many of the new AdS$_5$ vacua can also be found using more analytic methods. We generalize  here an observation in \cite{Bobev:2019dik} about the existence of a special truncation in four dimensions in which the scalar manifold is a product of mutually commuting Poincar\'e disks. That truncation arises from the subalgebra $\su(1,1)^7\subset \frak e_{7(7)}$ and can be obtained by imposing a discrete $\ZZ_2^3$ symmetry on the scalar fields. This truncation is quite remarkable in that it is very easy to analyze analytically and yet its potential captures 25\% of the 194 known critical points. 

A natural question is whether there exists a similarly  marvelous  truncation for the $\cals N=8$ supergravity in five dimensions. We find that indeed it does and corresponds to the embedding 
$\frak o(1,1)^2\oplus \su(1,1)\subset \frak e_{6(6)}$ for which  the scalar coset   is a product of 6 simple commuting factors,
\be\label{10scalarmanifold}
{\cal M}_{(10)} \equiv {\rm O}(1,1) ^2  \times \Big(\f{\SU(1,1)}{\U(1)}\Big)^4~,
\ee
that is 2 half-lines and 4 Poincar\'e disks. 
In fact, there exist two different consistent truncations for which the full scalar potential has been already worked out in the literature. Both use a $\ZZ_2^3$ symmetry and have the same looking coset, but  preserve different amount of the $\SO(6)\times \SL(2,\RR)$ symmetry. The first one, found 20 years ago in  \cite{Khavaev:2000gb}, has an $\rm U(1)^4$ unbroken symmetry so that the truncated potential can be reduced to  6 scalar fields. The second truncation,  found quite recently in  \cite{Bobev:2016nua}, breaks all continuous symmetries and the potential is a function of all 10 fields. For that reason we will refer to them as the 6-scalar and the 10-scalar model, respectively. As we show in Section~\ref{sec:truncation}, all critical points in the 6-scalar model lie  within the 10-scalar model. We find that the latter has 15 AdS$_4$ vacua, with many of the new critical points  computable  exactly and a few remaining ones easily accessible to standard numerical routines for example in  Mathematica. 

The 6-scalar and the  10-scalar models arise by imposing different $\ZZ_2$ symmetry on the same intermediate truncation of the $\cals N=8$ supergravity with respect to   $\ZZ_2^2\subset \SO(6)$. That truncation has the scalar coset,
\begin{equation}\label{18sccos}
\cals M_{(18)}\equiv {\rm O(1,1)^2\times {SO(4,4)\over S(O(4)\times O(4))}}\,,
\end{equation}
and preserves $\rm U(1)^4$ continuous symmetry. This means that the potential in this model depends on $18-4=14$ scalar fields. By carefully choosing the parametrization of both factors, we are able to compute the potential in a closed analytic form and determine, once more using  a simple numerical routine, that it has in total 17 critical points. 
This shows that  more than 50\% of all critical points are accessible analytically either exactly or by using a high level numerical routines to solve a system of explicit equations for the extrema of the potential. 

{An even stronger validation of the numerical results of  the TensorFlow search can be obtained by employing the so-called solvable parametrization \cite{Andrianopoli:1996bq,Andrianopoli:1996zg} of the full scalar coset. Indeed, we find that in this theory, as opposed to the ones in three or four dimensions, it is possible to compute the potential as an explicit function of all 42 scalar fields and then perform a search for critical points using Mathematica routines.  This parametrizaton also allows us to consider  systematically  the $\ZZ_2$-invariant truncations that encompass the analytic results above. In particular, it turns out that all 32 critical points can be found within a 22-dimensional subspace of the coset manifold that is invariant under the special $\ZZ_2$ symmetry used to arrive at the 10-scalar model. }

We begin in Section~\ref{sec:truncation} with a detailed discussion of the 10-scalar model and compute analytically whenever possible its 15 critical points.  In Section~\ref{sec:TFAdS} we describe the numerical search performed with TensorFlow and elucidate relevant differences between the computational strategies in the $d=5$ search here and the $d=4$ search in \cite{Comsa:2019rcz}. {In Section~\ref{sec:solvable}, we reproduce all critical points using the solvable parametrzation and summarize various $\ZZ_2$-invariant truncations.} 
We conclude with some open questions in Section~\ref{sec:conclusions}. A lot of technical details can be found in the appendices. Throughout the paper we use the same conventions as in \cite{Gunaydin:1985cu}. However, to avoid any ambiguities, in Appendix~\ref{appConv} we summarize the details of our parametrization of the scalar coset of the $\SO(6)$ supergravity. 
 Appendix~\ref{appendixA} has  a detailed discussion of the consistent truncation to 18 scalars in \eqref{18sccos}. We present a careful derivation of the full potential in this 14-scalars model and find its critical points. 
The results of the full numerical search can be found in Appendix~\ref{appendixB}. We give  a list of all 31 critical points together with their locations (partially canonicalized)  and the mass spectra. Finally, in Appendix~\ref{appendixC}, we collect some old results for the scalar mass spectra around the critical points in Table~\ref{tbl:KPW}, most of which where known to many but never published.

\medskip
\noindent
\textit{Note added in version 1:} The results in this paper were reported in seminars and at a conference  \cite{FTtalks2020,KPtalks2020}. While we were preparing this manuscript, we became aware of the recent work \cite{Krishnan:2020sfg}, which finds 32 critical points of the  scalar potential in  $\mathcal{N}=8$ $d=5$ gauged supergravity using TensorFlow. The authors of \cite{Krishnan:2020sfg} also calculate the gravitini spectra and find no new supersymmetric points. Apart from solution \#26 in \cite{Krishnan:2020sfg}, which is missing from our list, we find a complete  match between their values of the potential at critical points and the ones in our search. This provides yet another validation of the two numerical searches.
\medskip

\noindent
\textit{Note added:} Since the posting of the original version of this paper, we found all 32 critical points using our TensorFlow code that will soon be open sourced in the Google Research M-theory repository.\footnote{\href{https://github.com/google-research/google-research/tree/master/m_theory}{https://github.com/google-research/google-research/tree/master/m\_theory}.} These results are also independently confirmed by the new search using a combination of analytic and numerical routines in Mathematica based on the solvable parametrization of the potential as summarized in the newly added Section 4.

\section{The 10-scalar model}
\label{sec:truncation}

As explained in the Introduction, one way to deal with the complexity of the five-dimensional $\SO(6)$ gauged supergravity is to look for consistent truncations of the theory by imposing invariance under a subgroup, $G$, of the full symmetry group. In most examples $G\subset \SO(6)$, or $\SO(6)\times \SL(2,\RR)$, but the most interesting truncation discussed in this section is when $G\subset {\rm S(O(6)\times GL(2,\RR))}$.
Several such consistent truncations have been studied in the literature in various contexts, see for example \cite{Girardello:1998pd,Distler:1998gb,Girardello:1999bd,Pilch:2000ue,Pilch:2000fu,Bobev:2010de,Aprile:2011uq,Bobev:2013cja,Bobev:2014jva}, but no new critical points have been found after a 
systematic search within an $\SU(2)$ invariant truncation in~\cite{Khavaev:1998fb}. Other $\SU(2)$-invariant truncations listed in Table A.1 in \cite{Freedman:1999gp} merely reproduce a subset of critical points in  \cite{Khavaev:1998fb}. The same is true for an $\rm U(1)$-invariant truncation in \cite{Bobev:2014jva}. There are, however, two truncations with respect to discrete symmetries obtained in \cite{Khavaev:2000gb} and \cite{Bobev:2016nua}, respectively,  in which   ``holomorphic'' superpotentials,  and thus the full scalar potentials, are known explicitly. It appears that those potentials have never been fully analyzed. Therefore we begin our search by carefully examining these two models.

\subsection{The consistent truncation}
\label{subsec:trunc}

Motivated by this we start our discussion by studying the critical points of the consistent truncation in \cite{Bobev:2016nua} which is invariant under a $\mathbb{Z}^3$ subgroup of $\rm S(O(6)\times GL(2,\RR))$ and contains 10 out of the 42 scalar fields of the maximal theory. The procedure to obtain this truncation is outlined in detail in \cite{Bobev:2016nua} and so we will be brief. Consider the $\rm O(6)$ matrices
\be\label{Pis}
\begin{split}
	P_1 &= \text{diag}(-1,-1,1,1,1,1)~,\\ 
	P_2 &= \text{diag}(1,1,-1,-1,1,1)~,\\
	P_3 &= \text{diag}(1,-1,1,-1,1,-1)~,
\end{split}
\ee
and the following $\GL(2)$ matrices
\be\label{Qis}
\begin{split}
	Q  &= \text{diag}(-1,-1)~,\qquad 
	Q'  = \text{diag}(-1,1)~.
\end{split}
\ee
The truncation consists of the five-dimensional metric in addition to all fields that are invariant under the action of $P_1Q$, $P_2Q$,  $P_3Q'$. Even though the third matrix $P_3Q'$ is not inside $\SL(6,\RR)\times \SL(2,\RR)$, it is still a valid discrete symmetry to impose as explained in \cite{Pilch:2000fu,Bobev:2016nua}. In this paper we focus on AdS$_5$ vacua of the theory and so are only concerned with scalar fields that are invariant. Form fields must be set to zero. Imposing these symmetries leaves ten scalars parametrizing the scalar manifold
\be\label{10scalarmanifold}
{\cal M}_{(10)} = {\rm O}(1,1) \times {\rm O}(1,1)  \times \Big(\f{\SU(1,1)}{\U(1)}\Big)^4~.
\ee
These consist of five ${\bf 20}'$ scalars, four ${\bf 10}\oplus\overline{\bf 10}$ scalars, and the dilaton. Two of the ${\bf 20}'$ scalars are singled out as they parametrize the two ${\rm O}(1,1)$-factors in \eqref{10scalarmanifold}. An explicit parametrization of the coset~\eqref{10scalarmanifold} is given by specifying the generators of $\mathfrak{e}_{6(6)}$ that are invariant with respect to our choice of discrete symmetries. As explained in Appendix~\ref{appConv} we use an $\SL(6,\mathbb{R})\times\SL(2,\mathbb{R})$ basis to define our generators. In particular the two ${\rm O}(1,1) $-factors correspond to the generators $\mathfrak{g}_{\alpha}$ and $\mathfrak{g}_{\beta}$ defined by
\begin{equation}\label{galbe}
\begin{split}
\mathfrak{g}_\alpha & \eql \widehat\Lambda{} ^1{}_1+\widehat\Lambda{} ^2{}_2-\widehat\Lambda{} ^3{}_3-\widehat\Lambda{} ^4{}_4\,,\\
\mathfrak{g}_\beta & \eql \widehat\Lambda{} ^1{}_1+\widehat\Lambda{} ^2{}_2+\widehat\Lambda{} ^3{}_3+\widehat\Lambda{} ^4{}_4-2 \widehat\Lambda{} ^5{}_5-2 \widehat\Lambda{} ^6{}_6\,,
\end{split}
\end{equation}
using the notation in Appendix~\ref{appConv}.
The remaining scalars are best parametrized in terms of one of the non-compact generators of $\su(1,1)$ together with the compact one. The remaining non-compact generator can be obtained as the commutator of the other two. Using the notation in Appendix~\ref{appConv} the four compact generators $\mathfrak{r}$ and the non-compact generators $\mathfrak{t}$ are specified by
\be\label{trgenerators}
\begin{split}
\mathfrak{t}_1&= \frac{1}{\sqrt{2}}(\widehat\Sigma_{1357}-\widehat\Sigma_{2468})~,\qquad \mathfrak{r}_1= \frac{1}{\sqrt{2}}(\widehat\Sigma_{1357}+\widehat\Sigma_{2468})~,\\
\mathfrak{t}_2&= \frac{1}{\sqrt{2}}(\widehat\Sigma_{2367}-\widehat\Sigma_{1458})~,\qquad \mathfrak{r}_2= -\frac{1}{\sqrt{2}}(\widehat\Sigma_{2367}+\widehat\Sigma_{1458})~,\\
\mathfrak{t}_3&= \frac{1}{\sqrt{2}}(\widehat\Sigma_{2457}-\widehat\Sigma_{1368})~,\qquad \mathfrak{r}_3= -\frac{1}{\sqrt{2}}(\widehat\Sigma_{2457}+\widehat\Sigma_{1368})~,\\
\mathfrak{t}_4&= \frac{1}{\sqrt{2}}(\widehat\Sigma_{1467}-\widehat\Sigma_{2358})~,\qquad \mathfrak{r}_4= \frac{1}{\sqrt{2}}(\widehat\Sigma_{1467}+\widehat\Sigma_{2358})~.
\end{split}
\ee
The full ${\rm E}_{6(6)}$ group element is constructed as follows
\be\label{eq:cV10}
\mathcal{V} = \exp({\alpha \mathfrak{g}_\alpha})\cdot \exp({\beta {\mathfrak{g}}_\beta})\cdot \prod_{i=1}^4\exp(-{\omega_i \mathfrak{r}_i})\cdot\exp({\rho_i \mathfrak{t}_i})\cdot \exp({\omega_i \mathfrak{r}_i})~.
\ee
Notice that the commutator of the $\SU(1,1)$ generators $\mathfrak{t}_i$ and $\mathfrak{r}_i$ gives a linear combination of ${\bf 20}'$ generators $\widehat\Lambda{} ^I{}_J$ in addition to the dilaton generator $\mathfrak{g}_\text{dilaton} = \widehat\Lambda{}^7{}_7-\widehat\Lambda{} ^8{}_8$. It thus follows that one of the ten scalars is the dilaton. Since the scalar potential of the full $\SO(6)$ gauged theory does not depend on the dilaton, the same will be true in the truncated 10-scalar model. The way we have parametrized the manifold, $\cals M_{(10)}$, in \eqref{eq:cV10},
the dilaton is mixed with all the other $\SU(1,1)$ scalars and isolating it is difficult. The action of $\SL(2,\RR)$ on $\mathcal{V}$ is given by the transformation
\be
\mathcal{V}\mapsto \mathcal{V}\cdot \exp({t \,\mathfrak{g}_\text{dilaton}})~,
\ee
which leaves the potential invariant. Even though in principle it should be possible to translate what this action implies for the scalars $\rho_i$ and $\omega_i$, in practice the transformation is a complicated simultaneous action on all eight fields. 

The scalar potential of this truncation can be compactly written as
\be\label{eq:cP10}
{\cal P} = \frac1{32} \e^{\cal K}\left(\f{1}{6}| \partial_\alpha {\cal W}|^2 + \f{1}{2}|\partial_\beta {\cal W}|^2 + {\cal K}^{i\bar\jmath}D_i{\cal W} D_{\bar\jmath}\overline{\cal W} - \f83 |{\cal W}|^2\right)~,
\ee
where the K\"ahler covariant derivative is $D_i\mathcal{F} \equiv \partial_{i}\mathcal{F}+\mathcal{F}\partial_{i}\mathcal{K}$, the K{\"a}hler potential is
\be
{\cal K} = -\sum_{i=1}^4 \log(1-|z_i|^2)~,
\ee
and determines the kinetic terms through the K\"ahler metric $\mathcal{K}_{i\bar{\jmath}} \equiv \frac{\partial\mathcal{K}}{\partial z^{i}\partial \bar{z}^{
\bar{\jmath}}}$ and its inverse $\mathcal{K}^{i\bar{\jmath}}$. The superpotential is \cite{Bobev:2016nua}
\be
\begin{split}
{\cal W} = \e^{-4\alpha}(1+ z_1z_2 -z_1z_3-z_1z_4-z_2z_3-z_2z_4+z_3z_4+z_1z_2z_3z_4)& \\
+ \e^{2\alpha+2\beta}(1+ z_1z_2 +z_1z_3+z_1z_4+z_2z_3+z_2z_4+z_3z_4+z_1z_2z_3z_4)&\\
 +\e^{2\alpha-2\beta}(1- z_1z_2 + z_1z_3-z_1z_4-z_2z_3+z_2z_4-z_3z_4+z_1z_2z_3z_4)&~.
\end{split}
\ee
The complex scalars $z_i$ are related to the $\rho_i$ and $\omega_i$ in \eqref{eq:cV10} as follows:
\be
z_j = i \tanh\f{\rho_j}{2}~\e^{-i\omega_j}~.
\ee
The 10-scalar model exhibits a number of discrete symmetries, some of which were identified in \cite{Bobev:2016nua}. For example $z_i \mapsto \pm\bar z_i$ and $z_i\mapsto -z_i$. Here we would like to point out a rather large group of symmetries that leaves the superpotential invariant. It can be specified by
\begin{equation}
\begin{split}
e_1~&:~ \alpha \mapsto -\f{\alpha+\beta}2 ~,\quad \beta \mapsto \f{-3\alpha+\beta}2 ~,\quad z_1\mapsto -z_2 \mapsto z_1~.\\
e_2~&:~ \alpha \mapsto -\f{\alpha+\beta}2 ~,\quad \beta \mapsto \f{3\alpha-\beta}2 ~,\quad z_2\mapsto z_3 \mapsto -z_4\mapsto z_2~.\\
e_3~&:~ \alpha \mapsto \f{-\alpha+\beta}2 ~,\quad \beta \mapsto \f{3\alpha+\beta}2 ~,\quad z_1\mapsto -z_2\mapsto z_4\mapsto -z_3\mapsto z_1~.\\
\end{split}
\end{equation}
These satisfy $e_1^2 = e_2^3 = e_3^4=e_1e_2e_3 =1$ and therefore generate the group $S_4$. As we show in the next section  this model has 15 critical points including all five of \cite{Khavaev:1998fb}. Furthermore, by computing the masses of the 10 scalar fields we have checked that all non-supersymmetric critical points are perturbatively unstable within the 10-scalar model.

We note that a simpler six-scalar model can be obtained by setting the real parts of all $z_i$ scalars to zero. The potential then reduces to that of \cite{Khavaev:2000gb}, see also \cite{Bobev:2010de}. In \cite{Khavaev:2000gb} a 10-scalar truncation was considered which is different from the one we have been discussing here. This latter truncation has a potential which only depends on six fields. Explicitly, the potential for this 6-scalar model can be obtained from the one in \eqref{eq:cP10} by setting:
\be
\begin{split}
z_1 &= i\tanh \f12(\varphi_1-\varphi_2-\varphi_3+\varphi_4)~,\\
z_2 &= i\tanh \f12(\varphi_1+\varphi_2-\varphi_3-\varphi_4)~,\\
z_3 &= i\tanh \f12(\varphi_1+\varphi_2+\varphi_3+\varphi_4)~,\\
z_4 &= i\tanh \f12(\varphi_1-\varphi_2+\varphi_3-\varphi_4)~,
\end{split}
\ee
The potential can also be written in terms of a superpotential,
\be\label{eq:cP6}
{\cal P} = \f18\left( \f{1}{6}( \partial_\alpha W)^2 + \f{1}{2}(\partial_\beta W)^2 + (\partial_i W)^2- \f83 W^2\right)~,
\ee
where $\partial_i$ denotes a derivative with respect to $\varphi_i$ and \cite{Khavaev:2000gb}
\be\label{Wkw}
\begin{split}
W = \frac14\e^{-4\alpha}(+\cosh 2\varphi_1-\cosh 2\varphi_2-\cosh 2\varphi_3-\cosh 2\varphi_4)& \\
+ \frac14\e^{2\alpha+2\beta}(-\cosh 2\varphi_1-\cosh 2\varphi_2+\cosh 2\varphi_3-\cosh 2\varphi_4)&\\
 +\frac14\e^{2\alpha-2\beta}(-\cosh 2\varphi_1+\cosh 2\varphi_2-\cosh 2\varphi_3-\cosh 2\varphi_4)&~.
\end{split}
\ee
Four of the points in  Table~\ref{tbl:KPW} are critical points of the 6-scalar potential in \eqref{eq:cP6}. Only the $\SO(5)$ invariant point, $\tt T0780031$, lies outside it. There are in total eight critical points in the 6-scalar model.

\subsection{Critical points}
\label{subsec:10scpts}

Here we provide a list of the 15 critical points of the 10-scalar model potential in \eqref{eq:cP10}. Some of these can be obtained analytically. For the others we have used the {$\tt FindRoot[~\cdot~]$} routine in Mathematica.

\paragraph{T0750000 \cite{Gunaydin:1984qu,Gunaydin:1985cu,Pernici:1985ju}}
\be
z_a = 0\,,\quad \alpha=\beta=0\,.
\ee
\be
{\cal P } = -\f34=-0.750000\,.
\ee
Symmetry: $\rm SO(6)\,,\quad \cals N=8$.\\
Comment: Critical point of the 6-scalar model.

\paragraph{T0780031 \cite{Gunaydin:1985cu}}
\be
z_2=z_3=0~,\quad z_1=z_4=2-\sqrt{3}~,\quad \beta = 3\alpha = \f{\log 3}{8}\,.
\ee
\be
{\cal P } =-\f{3\times 3^{2/3}}{8}\,.
\ee
Symmetry: $\SO(5)\,,\quad \cals N=0$

\paragraph{T0839947 \cite{Khavaev:1998fb}}
\be
z_1=i(\sqrt{3}-2)\,, \quad z_2=z_3=-z_4=-z_1\,, \quad \quad 3\alpha = \beta = \f14\log 2\,.
\ee
\be
{\cal P } = -\f{2\times 2^{1/3}}{3} \approx -0.839947\,.
\ee
Symmetry: $\SU(2)\times\U(1)\,,\quad \cals N=2$.\\
Comment: Critical point of the 6-scalar model.
\paragraph{T0843750 \cite{Gunaydin:1985cu}}
\be
z_1=- i \sqrt{5-2\sqrt{6}}\,, \quad z_2=z_4=-z_3=-z_1\,, \quad \quad \alpha = \beta = 0\,.
\ee
\be
{\cal P } =-\f{27}{32}=-0.843750\,.
\ee
Symmetry: $\SU(3)\,,\quad \cals N=0$\\
Comment: Critical point of the 6-scalar model.

\paragraph{T0870297 \cite{Khavaev:1998fb}}
\be
z_1=z_2=0\,, \quad z_3=-z_4 = i\, \frac{2\sqrt{10}+\sqrt{15}-5}{2\sqrt{10}+\sqrt{15}+5} \quad \quad  \beta=-3\alpha = -\frac{1}{8}\log 10\,.
\ee
\be
{\cal P } = -\f{3\times 5^{2/3}}{8\times 2^{1/3}}~.
\ee
Symmetry: $\SU(2)\times \U(1)\,,\quad \cals N=0$\\
Comment: Critical point of the 6-scalar model. In \cite{Khavaev:1998fb}, see Table~\ref{tbl:KPW}, the symmetry of this point is listed  as $\SU(2)\times \U(1)^2$. The second $\U(1)$ factor is the compact generator of $\SL(2,\mathbb{R})$ which lies outside of the $\SO(6)$ gauge algebra. 


\paragraph{T0878939}
This point has $\mathcal{N}=0$ and is a critical point of the 6-scalar model found by setting
\be
z_1=z_2 =0\,,\quad z_3=-z_4 = i\frac{\sqrt{-X^2-(1-Y)^4Y^4+2X Y^2(1+6Y+Y^2)}}{X+(1-Y)^2Y^2+2\sqrt{X}Y(1+Y)}\,,
\ee
with
\be
\alpha = \frac{1}{24}\log X~,\quad \beta = \frac{1}{2}\log Y -\frac{1}{8}\log X\,.
\ee
The potential is then
\be\label{eq:PotXY3}
{\cal P } = - \f{X^2 + Y^4(1-Y^2)^2-2XY^2(3+4Y+3Y^2)}{16X^{1/3}Y^2(X+Y^2(1-Y)^2)}~.
\ee
This potential has two critical points, \T0878939 and \T1001482 which are correlated as follows. First one takes one of the four real roots of the equation
\be\label{eq:YpolXY3}
8-44 Y + 33 Y^2 + 74 Y^3 + 33 Y^4 - 44 Y^5 + 8 Y^6=0\,.
\ee
Note that the equation is self-reciprocal or palindromic and therefore the solutions come in inverse pairs which lead to the same cosmological constant. Use a given solution for $Y$ to find $X$ as a solution of the equation
\be\label{eq:XpolXY3}
5 X^2 +(1-Y)^4Y^4+2XY^2(1-10Y+Y^2)=0\,.
\ee
where the solution for $X$ must be correlated with the solution of $Y$. That is the choice of sign in the above second order equation for $X$ is correlated with which of the two different solution we start with for $Y$. For \T0878939 we then find the approximate values
\be
X= 0.006865\,, \quad Y = 0.283702\,, \quad \mathcal{P} =-0.878939\,.
\ee
The value of $\mathcal{P}$ can be obtained as a root of the polynomial
\be
729 + 5723136 \,{\cal P}^3 + 14123008 \,{\cal P}^6 + 8388608 \,{\cal P}^9\,.
\ee
Note that the \T0870298 is another critical point of \eqref{eq:PotXY3} with $X=10$, $Y=1$.

\paragraph{T0887636}
This point has $\mathcal{N}=0$ and is a critical point of the 6-scalar model located at
\be
\begin{split}
z_2&=-z_3=i\,\frac{1-\sqrt{Y+\sqrt{Y^2-1}}}{1+\sqrt{Y+\sqrt{Y^2-1}}}\,,\quad \beta=3\alpha= \frac{1}{8}\log\left(\frac{20+4\sqrt{34}}{3}\right)\,,\\
z_1 &= i\,\frac{X+\sqrt{X^2-1}-\sqrt{Y+\sqrt{Y^2-1}}}{X+\sqrt{X^2-1}+\sqrt{Y+\sqrt{Y^2-1}}}\,,\\
z_4 &= i\,\frac{(X+\sqrt{X^2-1})\sqrt{Y+\sqrt{Y^2-1}}-1}{(X+\sqrt{X^2-1})\sqrt{Y+\sqrt{Y^2-1}}+1}\,,
\end{split}
\ee
where the potential reduces to
\be\label{pot887636}
{\cal P } = \f{(1+Y)^{1/3}\Big(2(1+Y)(Y^2-3) - X^2(7+3Y)\Big)}{16\times 2^{2/3} X^{4/3}}\,,
\ee
and
\be
X^2 = \f{1}{243}(88+40\sqrt{34})~,\quad Y = \f19(-1+2\sqrt{34})\,.
\ee
The value of the potential is
\be
{\cal P } = -\frac{(196079 + 33524\sqrt{34})^{1/3}}{ 3^{7/3}\times  2^{8/3}}{}\approx -0.887636\,,
\ee
which is the smaller of the two real roots of the polynomial
\be\label{eq:PpolXY3}
107811+100392448 {\cal P }^3 + 143327232 {\cal P }^6\,.
\ee

\paragraph{T0892913}
\be
z_3 = i\frac{1}{\sqrt{5}}\,,\quad z_1=z_2=z_4=0\,, \quad 3\alpha = \beta = \f12\log 2~\,.
\ee
\be
{\cal P } = -\f{9}{8\times 2^{1/3}}\,.
\ee
Symmetry: $\cals N=0$\\
Comment: Critical point of the 6-scalar model.

\paragraph{T0964525}
\begin{equation}
\begin{split}
\alpha &= -0.0262713\,, ~~~ \beta= 0.254756\,,~~~ z_1= 0.224701 - i\, 0.487424\,,\\
z_2 &= 0.0709794\,, ~~~ z_3= -0.256605\,,~~  z_4=0.0116728 - i\, 0.507927\,.
\end{split}
\end{equation}
\begin{equation}
\mathcal{P} \approx - 0.9645259\,.
\end{equation}
Symmetry: $\cals N=0$

\paragraph{T0982778}
\be
z_1=-z_4=\frac{\sqrt{19-4\sqrt{22}}}{\sqrt{3}}~,\quad z_2=z_3=0.280116 +i\, 0.485175~,\quad \beta = 3\alpha = \frac{1}{8}\log (3/2)\,.
\ee
\be
{\cal P } =-\f{3\times3^{2/3}}{4\times 2^{2/3}}\,.
\ee
Symmetry: $\cals N=0$.\\
Comment: The values of $z_1$ and $z_2$ can be obtained as   roots of the polynomials $3-38X^2 +3 X^4$ and $4 + 14Y^2 +45 Y^4+14 Y^6+4 Y^8$, respectively.


\paragraph{T1001482}
This point has $\mathcal{N}=0$ and is a critical point of the 6-scalar model obtained in the same way as \T0878939 with the following approximate values for the roots of the polynomials in \eqref{eq:YpolXY3} and \eqref{eq:XpolXY3}
\be
X= 0.097733\,, \quad Y=0.337328\,, \quad \mathcal{P}=-1.001482\,.
\ee
Note that $\mathcal{P}$ is a root of the polynomial in \eqref{eq:PpolXY3}.

\paragraph{T1125000}
This point has $\mathcal{N}=0$ and is located at
\be
\begin{split}
&z_1=-z_3=-\sqrt{1+2Y-Y^2}+iY~,\quad z_2=-\bar{z}_4 = \f{1-X-z_1(1+X)}{1+X-z_1(1-X)}~,\\
&\beta =\frac{1}{2} \log X~,\quad \alpha=0~,
\end{split}
\ee
which gives the potential
\be
{\cal P} =-\f98~.
\ee
The value of $X$ is a root of the polynomial $1-5X^2 + X^4$. The value of $Y$ is unfixed. This is due to the fact that the five-dimensional dilaton is a flat direction in the potential. Therefore we can fix $Y$ to any convenient value by an $\SL(2,\RR)$ symmetry transformation. The only constraint when fixing $Y$ is that one has to ensure that all four scalars $z_a$ lie inside the unit disk.

\paragraph{T1304606}
\begin{equation}
\begin{split}
\alpha &= 0.0713344\,,\quad \beta= 0.214003\,,\\
z_1 &= 0.340985 - i\,0.385628 \,,\quad z_2 = 0.109181 +i\, 0.698203 \,,\\
z_3 &= 0.0805304 -i\,0.315369 \,,\quad  z_4=-0.481872 - i\,0.341603\,.
\end{split}
\end{equation}
\begin{equation}
\mathcal{P} \approx - 1.304606\,.
\end{equation}
Symmetry: $\cals N=0$

\paragraph{T1417411}
\be
z_1=-z_4=i\,\frac{\sqrt{(9-4\sqrt{2})(1+i\,4\sqrt{3})}}{7}\,, \quad z_2=z_3= i(\sqrt{2}-1)\,, \quad \beta = 3\alpha = \frac{1}{4}\log 2\,.
\ee
\be
{\cal P} =-\f9{4\times 2^{2/3}}\,.
\ee
Symmetry: $\cals N=0$

\paragraph{T1501862}
\begin{equation}
\begin{split}
\alpha &= 0.0766018\,,\quad \beta= 0.0519887\,,\\
z_1 &= -0.214941 +i\, 0.285334 \,,\quad z_2 = -0.0554356 + i\,0.297182  \,,\\
z_3 &= 0.483533 + i\, 0.610042 \,,  \quad ~~z_4=0.293764 - i\, 0.686 \,.
\end{split}
\end{equation}
\begin{equation}
\mathcal{P} \approx - 1.501862\,.
\end{equation}
Symmetry: $\cals N=0$

\section{Critical points with TensorFlow}
\label{sec:TFAdS}



\lstdefinestyle{pystyle}{
  basicstyle=\footnotesize
  }



Following the basic strategy explained in~\cite{Comsa:2019rcz}, the
numerical search for critical points was performed with TensorFlow.
In this section, we want to elucidate relevant differences between the
computational strategies used for~$d=4$ in this earlier publication
and~$d=5$ supergravity in this article. These mainly come from two
sources, differences in physics, and also advances in the software
ecosystem.

\subsection{TensorFlow and other options}

The commonly used conventions for de Wit-Nicolai maximal gauged~$d=4$
supergravity use complex~E$_{7(7)}$ generator matrices. When employing
numerical minimization with backpropagation as an effective strategy
to search for vacuum solutions of the equations of motion, the
stationarity condition is a smooth~$\mathbb{R}^{70}\to \mathbb{R}$
function. Using Machine Learning terminology, one would regard this as
the `Loss Function'. If we want to keep the code in close alignment
with the formulae from the published literature, we hence need a
framework for reverse-mode automatic differentiation (AD) that supports
Einstein summation, taking (ideally also higher) derivatives of matrix
exponentiation, complex matrix exponentiation, and, importantly,
taking gradients of $\mathbb{R}^{n}\to \mathbb{R}$ computations even
if intermediate steps involve complex quantities and holomorphic
functions.

It is especially this last point that is slightly subtle and
apparently not widely appreciated in the Machine Learning world, which
makes TensorFlow at the time of this writing (to the best of the
authors' knowledge) the only AD framework with which the~$d=4$
calculation could be done using the established conventions.
We want to briefly explain why.

For loss functions that involve complex intermediate quantities, it is
not sufficient for a computational framework to simply support complex
derivatives: it must be able to in particular correctly handle the
case that a real-valued result is the magnitude-square of a complex
intermediate result, schematically:
$y=f_j(z_k(x_m))\cdot\overline{f_j(z_k(x_m))}$, with the~$f_j$ being
holomorphic functions of the intermediate complex quantities~$z_k$
that in turn are functions of the real input parameters~$x_k$.  When
backpropagating such an expression, the AD framework repeatedly
answers the question by how much the final result would change,
relative to~$\varepsilon$, if one interrupted the calculation right
after the currently-in-focus intermediate quantity~$q_n$ was obtained
and changed it~$q_n\to q_n+\varepsilon$. This answer, i.e. the
sensitivity of the end result on~$q_n$, is found by referring, in
every step, to the already-known sensitivities for later intermediate
quantities~$q_{n+k}$. Starting with the sensitivity of the end result
on the end result, which is~$1$, we proceed through the entire
computation a second time, in reverse, to ultimately obtain the
sensitivities of the end result on the input parameters, i.e. the
gradient.  For a product of the above schematic form, the sensitivity
of the end result~$y$ on the intermediate quantity~$z_k(x_m)$
is~$\overline{f_j(z_k(x_m))}\cdot\partial_{z_k} f_j(z_k(x_m))$, and
the sensitivity of~$y$ on the intermediate
quantity~$\overline{z_k(x_m)}$ is the complex-conjugate of this value.
Clearly, a reverse mode automatic differentiation framework that only
knows about holomorphic derivatives and not this subtlety involving
complex conjugation will not be able to produce the expected
gradients. TensorFlow uses a modified definition of a `complex
gradient' that is \emph{not} the holomorphic derivative, but also
involves complex conjugation in precisely the way that is needed to
make this case work.\footnote{For
  technical details,
  cf. \href{https://github.com/tensorflow/tensorflow/issues/3348}{\texttt{https://github.com/tensorflow/tensorflow/issues/3348}}}

While the 56-dimensional fundamental representation of~E$_{7(-133)}$
is pseudoreal (i.e. does not permit all-real generator matrices), this
is not the case for~E$_{7(7)}$, closely paralleling the familiar
situation for~$\SU(2)$ and~$\SL(2,\RR)$. It is indeed possible to translate
de Wit-Nicolai supergravity from the `$\SU(8)$-aligned' basis that
makes fermion couplings look simple to a `$\SL(8,\RR)$-aligned' basis with
all-real~E$_{7(7)}$ generator matrices, and this alternative
description has been used e.g. in~\cite{DallAgata:2011aa}
to great effect. In maximal gauged five-dimensional supergravity,
the commonly used conventions employ a real basis for the
corresponding~E$_{6(6)}$ generator matrices of size~$27\times 27$,
and so there would be the option to also base the computation
on some other reverse-mode AD numerical framework, such as perhaps
the -- in comparison to TensorFlow -- much more lightweight `JAX'
library \cite{jax2018github}.

For this work, we nevertheless decided to stay with TensorFlow, partly
out of the desire to develop further software tools for supergravity
research that are generally applicable also in situations where
complex derivatives occur.

\subsection{The $d=5$ calculation}

As in maximal four-dimensional supergravity, critical points of the
equations of motion are saddle points, except for the 
maximum at the origin with unbroken~$\SO(6)$ symmetry.  For this work,
we did not use a stationarity condition that is expressed in terms of
the gradient of the potential with respect to an infinitesimal frame
change that multiplies the Vielbein matrix from one side, as in (2.8)
and (2.9) of~\cite{Comsa:2019rcz}. Rather, we took as stationarity
condition the length-squared of the gradient of the potential, and let
TensorFlow work out the gradient of this (scalar) stationarity
condition. The theory of Automated Differentiation tells us that the
computational effort for obtaining the gradient of a scalar function
is no more than six times the effort to compute the function (ignoring
the effect of caches), and so computing the gradient of the
stationarity-condition here is no more than~$6^2\times$ the effort
of evaluating the potential, which is quite affordable with
only~$42$ parameters.

For the de Wit-Nicolai theory, $\mathfrak{spin}(8)$ symmetry can be
employed to rotate a solution in such a way that one of the two
symmetric traceless matrices~$M_{\alpha\beta},
M_{\dot\alpha\dot\beta}$ that describe the location of a critical point
(cf. (D.3) in~\cite{Comsa:2019rcz}) gets diagonalized. For
five-dimensional maximal supergravity, we first performed a scan in
the full 42-dimensional parameter space, starting from~$100\,000$
seeded pseudorandom starting locations, and then checked that we
could indeed re-identify all solutions found in this way by
performing another (similarly large) scan using a reduced
coordinate-parametrization that set the non-diagonal entries
of the~$\Lambda^I{}_J$ and also the two~$\SL(2,\RR)/\SO(2)$ axion-dilaton
parameters to zero. As the volume of the~$\SO(6)$ orbit of
a solution is a function of the distance from the origin, one would
naturally expect these two different scanning methods to produce
any given solution with very different probability, and so
using only the latter, reduced, parametrization, might have
increased the risk of overlooking solutions.
Also, the conjecture that one can indeed always set the
axion-dilaton parameters to zero seems to be currently unproven.

As for the~$d=4$ calculation, we employed residual unbroken~$\SO(6)$
symmetry that is associated with degenerate entries on the diagonal
of~$\Lambda^I{}_J$ to further reduce the number of
non-zero~$\Sigma_{ijk;\alpha}$-coefficients, but there is no guarantee
in our tables that the number of parameters found in each case is
indeed minimal.

Given that TensorFlow currently is limited to performing calculations
with at most IEEE~754 64-bit float precision, and also the inherent
problems of solving nonlinear equation systems via minimization to
good accuracy, we found it effective to further increase the accuracy
of a solution-candidate as obtained from minimization via a modified
multi-dimensional Newton method. Here, one has to be careful due to
the presence of flat (``Goldstone mode'') directions in the potential
and hence also stationarity condition.

\subsection{Modern TensorFlow}

In this work, TensorFlow mostly serves as a ``fast numerical gradients''
library for high-dimensional numerical minimization. While it is useful to adopt
Machine Learning terminology for easier communication with other
(mostly Machine Learning) users of TensorFlow, this is not strictly
necessary. Due to the public release of TensorFlow2 in
September 2019,\footnote{Cf. \href{https://blog.tensorflow.org/2019/09/tensorflow-20-is-now-available.html}{\texttt{https://blog.tensorflow.org/2019/09/tensorflow-20-is-now-available.html}}}
which moves away from the explicit meta-programming paradigm,
much of the scaffolding that was used on the example Colab
notebook\footnote{\href{https://research.google.com/seedbank/seed/so_supergravity_extrema}{\texttt{https://research.google.com/seedbank/seed/so\_supergravity\_extrema}}}
published alongside~\cite{Comsa:2019rcz} can be eliminated.
In particular, the need for continuation-passing
techniques (such as provided by: {\tt call\_with\_critical\_point\_scanner()})
in order to evaluate a function ``in session context'' is now gone.

There broadly are two major approaches to reverse mode Automatic
Differentiation (AD), program-transformation based AD and tape-based
AD. TensorFlow1 was based on program transformation, where the
`program' is a description of a calculation in terms of a
(tensor-)arithmetic graph that can be evaluated on general purpose
CPUs or alternatively also hardware that is more specialized towards
parallel numerics, i.e. GPUs or Google's Tensor Processing
Units\footnote{Cf. \href{https://tinyurl.com/y6gmwfes}{\tt https://tinyurl.com/y6gmwfes}}
(TPUs). The Python programming language is here used as a
`Meta-Language' to manipulate `graph' objects that represent
computations.

TensorFlow2 tries to hide much of this meta-programming complexity by
making the graph invisible to the user and mostly following the
`tape-based' paradigm. Here, the idea is that the sequence of computational
steps in a calculation for which we want to have a fast and accurate gradient
are recorded on a `tape'. Once the calculation is done, the tape
is `played in reverse', in each step updating sensitivities of the
final result on intermediate quantities, in their natural
latest-to-earliest order. Pragmatically, this means that a TensorFlow2
`{\tt Tensor}' object can be seen as an envelope around a NumPy array
that can be tracked on a tape, but otherwise is passed around
and manipulated mostly like an array of numbers. This in particular
means that with TensorFlow2, interfacing with optimizers such
as {\tt scipy.optimize.fmin\_bfgs()} no longer requires a
TensorFlow-provided wrapper such as {\tt ScipyOptimizerInterface()},
or initiating numerical evaluation through an explicitly managed `session',
but instead can be done by simply wrapping up numpy-arrays in
TensorFlow tensors for gradient computations, roughly along these lines:
\\[2ex]

\lstset{style=pystyle}
\begin{lstlisting}[language=Python,basicstyle=\scriptsize]
def tf_minimize(tf_func, x0):
  """Minimizes a TensorFlow tf.Tensor -> tf.Tensor function."""
  def f_opt(xs):
    return tf_func(tf.constant(xs, dtype=tf.float64)).numpy()
  def fprime_opt(xs):
    t_xs = tf.constant(xs, dtype=tf.float64)
    tape = tf.GradientTape()
    with tape:
      tape.watch(t_xs)
      t_val = tf_func(t_xs)
    return tape.gradient(t_val, t_xs).numpy()
  opt = scipy.optimize.fmin_bfgs(
    f_opt, numpy.array(x0), fprime=fprime_opt, disp=0)
  return f_opt(opt), opt
\end{lstlisting}

\section{Critical points from a solvable parametrization}
\label{sec:solvable}

A solvable parametrization of the salar cosets in supergravity theories \cite{Andrianopoli:1996bq,Andrianopoli:1996zg} arises from  the Iwasawa decomposition \cite{helgason1979differential} of noncompact semisimple Lie groups, $G=KDN$, where $K$ is the maximal compact subroup, $D$ is a maximal ``noncompact torus'' and $N$ is a noncompact, nilpotent subgroup.
The scalar vielbein is then globally given by the group elements
\begin{equation}\label{scVsp}
\cals V\eql \exp\Big(\sum_{i=1}^\ell \varphi_i \,\frak h_i\Big)\exp\Big(\sum_{\alpha\in \Delta_+}\,x_\alpha \,\frak e_\alpha\Big)\,,
\end{equation}
where $\frak h_i$ are  generators of a noncompact Cartan subalgebra and $\frak e_\alpha$ are the  corresponding positive root generators. A clear advantage of this parametrization is that the first exponential of commuting generators is easy to compute, while the second one  collapses to a polynomial. 

In this section we summarize the results obtained by applying the solvable parametrization to the full scalar coset $\rm E_{6(6)}/USp(8)$ of the $\cals N=8$ $d=5$ supergravity.%
\footnote{In the context of this theory,  the solvable parametrization was first  employed in \cite{Bianchi:2000sm} to compute the full scalar potential in an $\SO(3)$-invariant truncation with a coset $\rm G_{2(2)}/SO(4)$.} It turns out that  the current computational capabilities of  Mathematica run on a  laptop suffice to obtain a closed form analytic expression for the full  potential as a function of all 42 scalar fields and then search numerically for its critical points. 

\subsection{Solvable parametrization}

The simplest choice for the noncompact Cartan subalgebra is to take the diagonal generators in $\sl(6,\RR)\times\sl(2,\RR)\subset \frak e_{6(6)}$,
\begin{equation}\label{}
\begin{split}
\frak h_1 & \eql {1\over \sqrt 2}(\widehat\Lambda{} ^1{}_1-\widehat\Lambda{} ^2{}_2)\,, \\
\frak h_2 & \eql {1\over\sqrt 6}(\widehat\Lambda{} ^1{}_1+\widehat\Lambda{} ^2{}_2-2 \widehat\Lambda{} ^3{}_3)\,,\\
\frak h_3 & \eql {1\over 2\sqrt 3}(\widehat\Lambda{} ^1{}_1+\widehat\Lambda{} ^2{}_2+\widehat\Lambda{} ^3{}_3-3 \widehat\Lambda{} ^4{}_4)\,,\\
\frak h_4 & \eql {1\over 2\sqrt 5}(\widehat\Lambda{} ^1{}_1+\widehat\Lambda{} ^2{}_2+\widehat\Lambda{} ^3{}_3+\widehat\Lambda{} ^4{}_4-4 \widehat\Lambda{} ^5{}_5)\,,\\
\frak h_5 & \eql {1\over \sqrt {30}}(\widehat\Lambda{} ^1{}_1+\widehat\Lambda{} ^2{}_2+\widehat\Lambda{} ^3{}_3+\widehat\Lambda{} ^4{}_4+\widehat\Lambda{} ^5{}_5-5 \widehat\Lambda{} ^6{}_6)\,,\\
\frak h_6 & \eql {1\over \sqrt 2}(\widehat\Lambda{} ^7{}_7-\widehat\Lambda{} ^8{}_8)\,,
\end{split}
\end{equation}
which are normalized such that $\Tr \frak h_i\frak h_j=6\,\delta_{ij}$. A natural set of the corresponding  positive and negative root generators, $\frak e_\alpha$ and $\frak f_\alpha$, respectively, is given in Table~\ref{tbl:roots}. We parametrize the positive roots, $\alpha\in\Delta_+$, in terms of their coordinates in the simple root basis,
\begin{equation}\label{rootnot}
[n_1n_2n_3n_4n_5n_6]\qquad \longleftrightarrow\qquad \alpha=\sum_{i=1}^6 n_i\alpha_i\,,
\end{equation}
where the simple roots, $\alpha_i$, are given explicitly by 
\begin{equation}\label{}
\begin{aligned}
\alpha_1 &\eql  \left(\sqrt{2},0,0,0,0,0\right)\,, &  
\alpha_2   & \eql \left(-\coeff{1}{\sqrt{2}},\sqrt{\coeff{3}{2}},0,0,0,0\right)\,,\\ 
\alpha_3   & \eql \left(0,-\sqrt{\coeff{2}
   {3}},\coeff{2}{\sqrt{3}},0,0,0\right)\,,&\qquad 
\alpha_4   &\eql \left(0,0,-\coeff{\sqrt{3}}{2},\coeff{\sqrt{5}}{2},0,0\right)\,,\\
\alpha_5   & \eql  \left(0,0,0,-\coeff{2}{\sqrt{5}},\sqrt{\coeff{6}{5}},0\right)\,,&    \alpha_6   &\eql  \left(0,0,-\coeff{\sqrt{3}}{2},-\coeff{3}{2
   \sqrt{5}},-\sqrt{\coeff{3}{10}},\coeff{1}{\sqrt{2}}\right)\,.
\end{aligned}
\end{equation}
The generators $\Xi^\pm_{IJK\alpha}$ are defined by
\begin{equation}\label{}
\Xi^+_{IJK\alpha}\equiv {1\over \sqrt 2}\,\widehat\Sigma^{IJK\alpha}\,,\qquad 
\Xi^-_{IJK\alpha}\equiv {1\over \sqrt 2}\,\widehat\Sigma_{IJK\alpha}\,.
\end{equation}
The normalization is chosen such that
\begin{equation}\label{}
\Tr \frak e_\alpha \frak e_\beta\eql \Tr \frak f_\alpha \frak f_\beta\eql 0\,,\qquad \Tr \frak e_\alpha \frak f_\beta\eql 6\,\delta_{\alpha\beta}\,. 
\end{equation}

\begin{table}[H]
\begin{center}
\resizebox{0.6\textwidth}{!}{%
\begin{tabular}{@{\extracolsep{30 pt}}lll ccc}
\toprule
$\alpha\in\Delta_+$ & $\frak e_\alpha$ & $\frak f_\alpha$  & $S_1$ & $S_2$ & $S_3$ \\
\noalign{\smallskip}
\midrule
\noalign{\smallskip}
 \text{[100000]} & $\widehat\Lambda{}^1{}_2$ & $\widehat\Lambda{}^2{}_1$  & $*$ & $*$ &\\
 \text{[010000]} & $\widehat \Lambda{} ^2{}_3$ & $\widehat\Lambda{}^3{}_2$ && \\
 \text{[001000]} & $\widehat\Lambda{}^3{}_4$ & $\widehat\Lambda{}^4{}_3$ &  $*$ & $*$ &\\
 \text{[000100]} & $\widehat\Lambda{}^4{}_5$ & $\widehat\Lambda{}^5{}_4$  & $*$ & \\
 \text{[000010]} & $\widehat\Lambda{}^5{}_6$ & $\widehat\Lambda{}^6{}_5$ & $*$ & $*$\\
 \text{[000001]} & $\Xi ^+_{4567}$ & $\Xi ^-_{4567}$ && $*$
   \\[6pt]
 \text{[110000]} & $\widehat\Lambda{}^1{}_3$ & $\widehat\Lambda{}^3{}_1$ &&&  $*$ \\
 \text{[011000]} & $\widehat\Lambda{}^2{}_4$ & $\widehat\Lambda{}^4{}_2$ &&& $*$ \\
 \text{[001100]} & $\widehat\Lambda{}^3{}_5$ & $\widehat\Lambda{}^5{}_3$ & $*$ &&  $*$ \\
 \text{[000110]} & $\widehat\Lambda{}^4{}_6$ & $\widehat\Lambda{}^6{}_4$ & $*$ && $*$ \\
 \text{[001001]} & $\Xi ^+_{3567}$ & $\Xi ^-_{3567}$ && $*$ & $*$
   \\[6 pt]
 \text{[111000]} & $\widehat\Lambda{}^1{}_4$ & $\widehat\Lambda{}^4{}_1$ && \\
 \text{[011100]} & $\widehat\Lambda{}^2{}_5$ & $\widehat\Lambda{}^5{}_2$ && $*$ & \\
 \text{[001110]} & $\widehat\Lambda{}^3{}_6$ & $\widehat\Lambda{}^6{}_3$ & $*$ \\
 \text{[011001]} & $\Xi ^+_{2567}$ & $\Xi ^-_{2567}$ & $*$
   \\
 \text{[001101]} & $\Xi ^+_{3467}$ & $\Xi ^-_{3467}$
   \\[6 pt]
 \text{[111100]} & $\widehat\Lambda{}^1{}_5$ & $\widehat\Lambda{}^5{}_1$ && $*$ & $*$\\
 \text{[011110]} & $\widehat\Lambda{}^2{}_6$ & $\widehat\Lambda{}^6{}_2$ && $*$ & $*$\\
 \text{[111001]} & $\Xi ^+_{1567}$ & $\Xi ^-_{1567}$ & $*$&& $*$
   \\
 \text{[011101]} & $\Xi ^+_{2467}$ & $\Xi ^-_{2467}$ & $*$ & $*$ & $*$
   \\
 \text{[001111]} & $\Xi ^+_{3457}$ & $\Xi ^-_{3457}$ &&& $*$
   \\[6 pt]
 \text{[111110]} & $\widehat\Lambda{}^1{}_6$ & $\widehat\Lambda{}^6{}_1$ && $*$\\
 \text{[111101]} & $\Xi ^+_{1467}$ & $\Xi ^-_{1467}$ & $*$ & $*$ 
   \\
 \text{[012101]} & $\Xi ^+_{2367}$ & $\Xi ^-_{2367}$ & $*$ & $*$
   \\
 \text{[011111]} & $\Xi ^+_{2457}$ & $\Xi ^-_{2457}$ & $*$ & $*$ 
   \\[6 pt]
 \text{[112101]} & $\Xi ^+_{1367}$ & $\Xi ^-_{1367}$ & $*$ & $*$ & $*$
   \\
 \text{[111111]} & $\Xi ^+_{1457}$ & $\Xi ^-_{1457}$ & $*$ & $*$ & $*$
   \\
 \text{[012111]} & $\Xi ^+_{2357}$ & $\Xi ^-_{2357}$ & $*$ & $*$ & $*$
   \\[6 pt]
 \text{[122101]} & $\Xi ^+_{1267}$ & $\Xi ^-_{1267}$
   \\
 \text{[112111]} & $\Xi ^+_{1357}$ & $\Xi ^-_{1357}$ & $*$ & $*$
   \\
 \text{[012211]} & $\Xi ^+_{2347}$ & $\Xi ^-_{2347}$ & $*$
   \\[6 pt]
 \text{[122111]} & $\Xi ^+_{1257}$ & $\Xi ^-_{1257}$ &&& $*$
   \\
 \text{[112211]} & $\Xi ^+_{1347}$ & $\Xi ^-_{1347}$ & $*$ && $*$
   \\[6 pt]
 \text{[122211]} & $\Xi ^+_{1247}$ & $\Xi ^-_{1247}$ && $*$
   \\[6 pt]
 \text{[123211]} & $\Xi ^+_{1237}$ & $\Xi ^-_{1237}$ && $*$ & $*$
   \\[6 pt]
 \text{[123212]} & $\widehat\Lambda{}^7{}_8$ & $\widehat\Lambda{}^8{}_7$ & $*$ & $*$ \\
 \noalign{\smallskip}
\bottomrule
\noalign{\smallskip}
\end{tabular}
}
\caption{\label{tbl:roots} Root generators of $\rm E_{6(6)}$.}
\end{center}
\label{default}
\end{table}%

One should note that the root generators, $\frak e_\alpha$ and $\frak f_\alpha$ are combinations of both  compact and noncompact generators of $\frak e_{6(6)}$. Specifically, the compact generators  are spanned by $(\frak e_\alpha-\frak f_\alpha)$, while the noncompact ones by $(\frak e_\alpha+\frak f_\alpha)$ and the Cartan generators, $\frak h_i$. As a result, the relation between the scalar fields in the symmetric gauge used in the previous sections and the solvable parametrization here is highly nonlinear.\footnote{At a given point on the scalar coset, the relation between the two sets of fields is easily determined, at least numerically,  from the $\rm USp(8)$-invariant product of the scalar vielbein and its transpose.} In particular, the action of the $\rm SO(6)$ gauge group, which is very simple in the symmetric gauge, becomes completely obscured in the solvable parametrization.  

\subsection{The scalar potential and critical points}

A direct evaluation of the exponentials for the scalar 27-bein in \eqref{scVsp} shows that the second factor is a polynomial of degree 18. After symbolic substitution for the nonvanishing matrix elements of the scalar vielbein, it turns out possible to generate a close form expression for the full scalar potential by following the usual steps in \cite{Gunaydin:1985cu}. The resulting analytic expression in terms of the 42 scalar fields, $\varphi_i$, $i=1,\ldots,6$, and $x_\alpha$, $\alpha\in\Delta_+$, is given in the 
ancillary file.\footnote{The scalar fields in the file are  $\tt \varphi[i]$ and $\tt x[n_1,\ldots,n_6]$, where $[n_1\ldots n_6]$ denotes the root as in \eqref{rootnot}.} One can check numerically that, as expected,  the potential does not depend on the $\SL(2,\RR)$ scalar corresponding to the maximal root $[123212]$. 

After evaluating symbolically the gradient of the potential with respect to all scalar fields, we have performed an exhaustive numerical search for the critical points  using the ${\tt FindRoot[~\cdot~]}$ routine in Mathematica starting at random points on the scalar coset. Since the solvable parametrization of the coset does not lead to any coordinate singularities, unlike the polar parametrization used in a similar numerical search in Appendix~\ref{appendixA}, all zeros of the gradient correspond to actual critical points of the potential. The resulting list of critical points found in this search is the same as the one found using TensorFlow in Section~\ref{sec:TFAdS} that are  given in Appendix~\ref{appendixB}. This provides a completely independent consistency check between the two searches within the numerical accuracy of the Mathematica routines.

\subsection{$\ZZ_2$ truncations}

Given an analytic expression for the full potential, it is now straightforward to explore various truncations to smaller sectors. In particular, truncations with respect to the $\ZZ_2$ discrete symmetries considered in Section~\ref{sec:truncation} and Appendix~\ref{appendixA} amount to setting various subsets of the scalar  fields to zero. This results in simpler potentials, whose critical points can be determined using the same routine as for the full potential above.

\begin{table}[H]
\begin{center}
\resizebox{0.75\textwidth}{!}{%
\begin{tabular}{@{\extracolsep{30 pt}}lll ccc}
\toprule
Point & $S_{1,2}$ & $S_3$  & $S_1S_2$ & $S_{1,2}S_3$ & $S_1S_2S_3$ \\
\noalign{\smallskip}
\midrule
\noalign{\smallskip}
\tt T0750000 & $*$ & $*$ & $*$ & $*$ & $*$ \\
\tt T0780031 & $*$ & $*$ & $*$ & $*$ & $*$  \\
\tt T0839947 & $*$ & $*$ & $*$ & $*$ & $*$ \\
\tt T0843750 & $*$ & $*$ & $*$ & $*$ & $*$ \\
\tt T0870297 & $*$ & $*$ & $*$ & $*$ & $*$ \\
\tt T0878939 & $*$ & $*$ & $*$ & $*$ & $*$ \\
\tt T0887636 & $*$ & $*$ & $*$ & $*$ & $*$ \\
\tt T0892913 & $*$ & $*$ & $*$ & $*$ & $*$ \\
\tt T0963952 &   & $*$ &   &   &   \\
\tt T0964097 & $*$ & $*$ &   & $*$ &   \\
\tt T0964525 & $*$ & $*$ & $*$ & $*$ & $*$ \\
\tt T0982778 & $*$ & $*$ & $*$ & $*$ & $*$ \\
\tt T1001482 & $*$ & $*$ & $*$ & $*$ & $*$ \\
\tt T1054687 & $*$ & $*$ & $*$ & $*$ &   \\
\tt T1073529 & $*$ & $*$ &   & $*$ &   \\
\tt T1125000 & $*$ & $*$ & $*$ & $*$ & $*$ \\
\tt T1297247 &   & $*$ &   &   &   \\
\tt T1302912 & $*$ & $*$ &   & $*$ &   \\
\tt T1304606 & $*$ & $*$ & $*$ & $*$ & $*$ \\
\tt T1319179 & $*$ & $*$ &   & $*$ &   \\
\tt T1382251 &  & $*$ &   &  &   \\
\tt T1391035 & $*$ & $*$ &   &   &   \\
\tt T1416746 & $*$ & $*$ &   &   &   \\
\tt T1417411 & $*$ & $*$ & $*$ & $*$ & $*$ \\
\tt T1460654 &   & $*$ &   &   &   \\
\tt T1460729 & $*$ & $*$ &   & $*$ &   \\
\tt T1497042 & $*$ & $*$ &   & $*$ &   \\
\tt T1499666 & $*$ & $*$ &   &   &   \\
\tt T1501862 & $*$ & $*$ & $*$ & $*$ & $*$ \\
\tt T1510900 & $*$ & $*$ &   & $*$ &   \\
\tt T1547778 & $*$ & $*$ & $*$ & $*$ &   \\
\tt T1738407 & $*$ & $*$ &   & $*$ &   \\
 \noalign{\smallskip}
 \midrule
 \noalign{\smallskip}
 Total & 28 & 32 & 17 & 25 & 15\\
  \noalign{\smallskip}
\bottomrule
\noalign{\medskip}
\end{tabular}
}
\caption{\label{tbl:crptstr} Critical points from discrete truncations.}
\end{center}
\label{default}
\end{table}%

The three $\ZZ_2$ symmetries we want to discuss here are generated by
\begin{equation}\label{}
S_1\equiv P_1Q\,,\qquad S_2\equiv P_2Q\,,\qquad S_3\equiv P_3Q'\,,
\end{equation}
where $P_1$,  $P_2$, $P_3$ and $Q$, $Q'$ are given in \eqref{Pis} and \eqref{Qis}, respectively. Clearly, the Cartan generators, $\frak h_i$, commute with these symmetries. 
The root generators that are even (invariant) under a given symmetry are labelled by the star in Table~\ref{tbl:roots}. The remaining generators are odd and the corresponding scalar fields are set to zero in the truncations.

The results of our searches for critical points in various $\ZZ_2$, $\ZZ_2^2$ and $\ZZ_2^3$-invariant sectors are summarized in Table~\ref{tbl:crptstr}. Since $S_1$ and $S_2$ are conjugate under the adjoint action of $\SO(6)$, the two $\ZZ_2$-invariant truncations yield the same set of points, albeit with different sets of   scalar fields. The truncation to the sector invariant under $S_1$ and $S_2$ reproduces the points found in Appendix~\ref{appendixA}, where we use a completly different parametrization of the coset. The combined truncation with respect to $S_1$, $S_2$ and $S_3$ yields the 10-scalar model and we reproduce the results in Section~\ref{sec:truncation}. 

What is surprising and new here is that {\it all\/} critical points are found within a truncation  with respect to the special $\ZZ_2$ symmetry generated by $S_3\in {\rm S(O(6)\times
 GL(2,\RR))}$. The 22 scalars in this truncation parametrize the coset
\begin{equation}\label{}
\cals M_{(22)}\equiv {{\SL(6,\RR)}\over \SO(6)}\times {\SL(2,\RR)\over \SO(2)}\,,
\end{equation}
with, however, different $\SL(6,\RR)$ and $\SL(2,\RR)$ than those generated by $\widehat\Lambda{}^I{}_J$ and $\widehat\Lambda{}^\alpha{}_\beta$ in \eqref{gener}. From the last column in Table~\ref{tbl:roots} we see that the relevant  $\sl(6,\RR)\oplus \sl(2,\RR)$ subalgebra of $\frak e_{6(6)}$ is spanned by the ``even'' root generators with $n_1+\ldots+n_6\in 2\,\ZZ$. It would be interesting to understand  the a priori reason for the ``critical efficiency'' of this truncation.

\section{Conclusions}
\label{sec:conclusions}

In this paper we presented a numerical exploration of the AdS$_5$ vacua corresponding to critical points of the scalar potential of the $\SO(6)$ maximal gauged supergravity. Out of the 31 critical points, we find that there are only 2 that are supersymmetric and perturbatively stable. Usually one would dismiss the 29 unstable AdS$_5$ solutions as physically irrelevant. Nevertheless, the existence of these critical points may point towards some interesting dynamics in the supersymmetry broken phases of the planar $\mathcal{N}=4$ SYM theory. Perhaps some of these vacua admit an interpretation as holographic duals to complex CFTs \cite{Gorbenko:2018ncu,Faedo:2019nxw} or can serve as lampposts for other type of approximately conformal QFT dynamics similar to the ones studied in \cite{Donos:2017sba}. To understand this question better one can study holographic RG flows represented by domain wall solutions connecting our new vacua. This can be done most explicitly for the 10-scalar and 6-scalar consistent truncations. For example, if there are supersymmetric RG flows that closely approach some of the unstable AdS$_5$ vacua this may suggest an approximately conformal supersymmetric phase of $\mathcal{N}=4$ SYM. It should also be noted that the 10- and 14-scalar consistent truncations have wider applications in the context of holography. As emphasized in \cite{Bobev:2016nua,Bobev:2019wnf} they can be used to study the holographic dual description of the $\mathcal{N}=1^{*}$ mass deformation of $\mathcal{N}=4$ SYM on $\mathbb{R}^4$ and $S^4$ for general values of the complex mass parameters. 

All of the AdS$_5$ vacua we constructed can be uplifted to solutions of type IIB supergravity using the explicit formulae in \cite{Baguet:2015sma}. This will result in ten-dimensional AdS$_5$ solutions with non-trivial fluxes on $S^5$. Given that the new critical points  are perturbatively unstable, they can be used as a test ground for exploring the general mechanisms responsible for instabilities in non-supersymmetric flux compactifications. In addition, using the ten-dimensional uplift may allow for the possibility of stabilizing some of the AdS$_5$ vacua by projecting out the unstable modes using an appropriate orbifold action in type IIB string theory \cite{Bobev:2010ib}.

Finally, we note that there are other gaugings that lead to maximal supergravity theories in five dimensions, see \cite{Gunaydin:1985cu,deWit:2002vt}. It will be interesting to apply similar numerical and analytical tools  to study the critical points of these theories.

\bigskip
\bigskip
\leftline{\bf Acknowledgements}
\smallskip
\noindent 
We  are grateful to Jesse van Muiden and  Nick Warner for interesting discussions. T.F. would like to thank Jyrki Alakuijala, George Toderici, Ashok Popat, and Rahul Sukthankar for encouragement and support, and Rasmus Larsen for providing expertise on low level TF internals.
 The work of NB is supported in part by an Odysseus grant G0F9516N from the FWO and by the KU Leuven C1 grant ZKD1118 C16/16/005. FFG is a Postdoctoral Fellow of the Research Foundation - Flanders (FWO). KP is supported in part by DOE grant DE-SC0011687. NB, FFG, and KP are grateful to the Mainz Institute for Theoretical Physics (MITP) of the DFG Cluster of Excellence PRISMA$^+$ (Project ID 39083149), for its hospitality and its partial support during the initial stages of this project. KP would like to thank the ITF at KU Leuven for hospitality during part of this work.



\appendix
\section{Conventions}
\label{appConv}

Throughout this paper we use the same conventions as in \cite{Gunaydin:1985cu}, which the reader should consult for  details. Here we  summarize an explicit parametrization of the scalar manifold \begin{equation}\label{}
\cals M_{(42)}\equiv \rm {E_{6(6)}\over USp(8)}\,,
\end{equation}
of the $\cals N=8$ $d=5$ supergravity as needed for the  truncations in Section~\ref{sec:truncation} and Appendix~\ref{appendixA}, an explicit construction of the potential in Section~\ref{sec:TFAdS}, and specifying the location of its  critical points in Appendix~\ref{appendixB}. 

The most straightforward description of the $\frak e_{6(6)}$  generators  in the so-called $\rm SL(6,\RR)\times SL(2,\RR)$  basis is through their action on 27-dimensional vectors with components $(z_{IJ},z^{I\alpha})$,  $z_{IJ}=-z_{JI}$,\footnote{For the corresponding $27\times 27$ matrix, see (A.36) in \cite{Gunaydin:1985cu}.}
\begin{equation}\label{delzs}
\begin{split}
\delta z_{IJ} & \eql -\Lambda^K{}_Iz_{KJ}-\Lambda^K{}_Jz_{IK}+\Sigma_{IJK\beta}z^{K\beta}\,,\\[6 pt]
\delta z^{I\alpha} & \eql \Lambda^I{}_Kz^{K\alpha}+\Lambda^\alpha{}_\beta z^{I\beta}+\Sigma^{KLI\alpha}z_{KL}\,,
\end{split}
\end{equation}
where $(\Lambda^I{}_J)$ and $(\Lambda^\alpha{}_\beta)$ are  real matrices in  $\sl(6,\RR)$ and $\sl(2,\RR)$, respectively, and  $\Sigma_{ IJK \alpha}=\Sigma_{[IJK]\alpha}$ is real with 
\begin{equation}\label{SigmaUp}
\Sigma^{IJK\alpha}={1\over 6}\epsilon^{IJKLMN}\epsilon^{\alpha\beta}\Sigma_{LMN\beta}\,. 
\end{equation}
Note that the transformation \eqref{delzs} can be extended to arbitrary $(\Lambda^I{}_J)\in\gl(6,\RR)$ and $(\Lambda^\alpha{}_\beta)\in \gl(2,\RR)$. This can be used to introduce a convenient basis  of generators  $ (\widehat\Lambda{}^I{}_J,\widehat\Lambda{}^\alpha{}_\beta ,\widehat\Sigma_{IJK\alpha})$ in $\frak e_{6(6)}\oplus\RR^2$ defined by the following nonvanishing  parameters in \eqref{delzs} for each 
generator:\footnote{Note that unlike \cite{Gunaydin:1985cu} we use the range $\alpha,\beta=7,8$ for the $\SL(2,\RR)$ indices.}
\begin{equation}\label{gener}
\begin{split}
\widehat\Lambda{}^I{}_J & :\qquad \Lambda^I{}_J\eql 1\,,\qquad I,J=1,\ldots,6\,,\\
\widehat\Lambda{}^\alpha{}_\beta & :\qquad \Lambda^\alpha{}_\beta\eql 1\,,\qquad \alpha,\beta=7,8\,,\\
\widehat\Sigma_{IJK\alpha} & :\qquad \Sigma_{IJK\alpha}=\Sigma_{KIJ\alpha}\eql\ldots\eql -\Sigma_{KJI\alpha}\eql  1\,,\qquad I<J<K\,.
\end{split}
\end{equation}

The   coset, $\rm E_{6(6)}/USp(8)$,  has a trivial topology of $\RR^{42}$ and, via the exponential map, is isomorphic to the corresponding quotient of the Lie algebras, $\frak e_{6(6)}/\frak{usp}(8)$. The usual choice of the coset representatives is then given by the noncompact generators for which  
\begin{equation}\label{noncomp}
\Lambda^I{}_J\eql \Lambda^J{}_I\,,\qquad \Lambda^\alpha{}_\beta\eql \Lambda^\beta{}_\alpha,\qquad \Sigma_{IJK\alpha}\eql\Sigma^{IJK\alpha}\,.
\end{equation}
An ordered set of the $20+2+20$ independent parameters in \eqref{noncomp} provides then global coordinates on the scalar manifold, $\cals M_{(42)}$.

\section{14-scalar model}
\label{appendixA}

In this appendix we present   a truncation of the potential to a 14-scalar model that arises as an intermediate step in the construction of the 6-scalar model in  \cite{Khavaev:2000gb} and/or the 10-scalar model \cite{Bobev:2016nua} discussed in Section~\ref{sec:truncation}. The main result is an explicit, albeit rather complicated, form of the scalar potential in this sector. It yields a subset of 17 extrema of the full potential. 

\subsection{$\ZZ_2\times \ZZ_2$-invariant truncations}
There are two equivalent methods to obtain the 14-scalar model we are interested in. The first one is to truncate with respect to a  $\ZZ_2\times\ZZ_2\subset \SO(6)$   symmetry
generated by  \cite{Khavaev:2000gb}
\begin{equation}\label{KWz2z2}
g_1\eql \mathop\text{diag}(-1,-1,-1,-1,1,1) \qquad \text{and}\qquad g_2\eql \mathop\text{diag}(1,1,-1,-1,-1,-1)\,.
\end{equation}
The second method is to use $\ZZ_2\times \ZZ_2\subset{\rm S(O(6)\times \GL(2,\RR))}$ generated by 
 $P_1Q$ and $P_2Q$ \cite{Bobev:2016nua}, where 
   \begin{equation}\label{BBz2z2}
P_1\eql \mathop\text{diag}(-1,-1,1,1,1,1)\,,\qquad P_2\eql \mathop\text{diag}(1,1,-1,-1,1,1)\,,\qquad Q\eql  \mathop\text{diag}(-1,-1)\,.
\end{equation}
are the same as in \eqref{Pis} and \eqref{Qis}.
The truncations with respect to   the  $\ZZ_2\times \ZZ_2$ in \eqref{KWz2z2} or \eqref{BBz2z2}, respectively,   yield the same set  of invariant generators of $\frak o(1,1)^2\times \so(4,4)\subset \frak e_{6(6)}$, with the resulting scalar coset 
\begin{equation}\label{}
\cals M_{\rm O(1,1)^2}\times \cals M_{\rm SO(4,4)}\equiv\rm O(1,1)^2\times {SO(4,4)\over SO(4)\times SO(4)}\,.
\end{equation}
To compute the potential, we need a workable parametrization of the second factor.

\subsection{Polar parametrization of the coset}

In the vector representation of $\SO(4,4)$, the compact $\rm S(O(4)\times O(4))$ subgroup is given by  block matrices
\begin{equation}\label{}
O\eql \left(\begin{matrix}
O_1 & 0 \\ 0  & O_2
\end{matrix}\right)\,,\qquad O_1,O_2\in {\rm O(4)}\,,\quad   O_1O_2\in \SO(4)\,.
\end{equation}
The non-compact generators are of the form
\begin{equation}\label{}
X\eql \left(\begin{matrix}
0 & M \\ M^T & 0
\end{matrix}\right)\,,
\end{equation}
and the  $4\times 4$ matrices, $M$, provide global coordinates on the coset. Now, note
that
\begin{equation}\label{}
OXO^T\eql \left(\begin{matrix}
0 & O_1MO_2^T \\ O_2M^TO_1^T & 0
\end{matrix}\right)\,,
\end{equation}
and use the fact that any generic real matrix can be diagonalized by two orthogonal matrices, that is
\begin{equation}\label{}
M\eql O_1\Lambda O_2^T\,.
\end{equation}
 The diagonal matrix, $\Lambda$, consists of 4 commuting, noncompact generators. Futhermore, any 4 such generators are conjugate under the action of the compact subgroup. The idea now is to parametrize the $\cals M_{\SO(4,4)}$ coset in terms of Euler angles for $O_1$ and $O_2$ and the four parameters in $\Lambda$. 
 
To this end, we first decompose the compact generators of $\so(4,4)\subset \frak e_{6(6)}$ into generators of 4 mutually commuting $\su(2)$'s, which are labelled by  $\alpha,\beta,\gamma,\delta$.\footnote{Note that $\alpha$ and $\beta$ have different meaning in the main text than in this appendix.} Inside the $\frak e_{6(6)}$, one can choose those generators as follows:
\begin{equation}\label{}
\begin{split}
\frak r_1^{(\alpha)} & \eql {1\over \sqrt{2}}\left(-\widehat\Sigma _{1357}+\widehat\Sigma _{1368}+\widehat\Sigma _{1458}+\widehat\Sigma _{1467}+\widehat\Sigma _{2358}+\widehat\Sigma _{2367}+\widehat\Sigma _{2457}-\widehat\Sigma _{2468}\right)\,,\\
\frak r_2^{(\alpha)} &\eql{1\over \sqrt{2}}\left( -\widehat\Sigma _{1358}-\widehat\Sigma _{1367}-\widehat\Sigma _{1457}+\widehat\Sigma _{1468}-\widehat\Sigma _{2357}+\widehat\Sigma _{2368}+\widehat\Sigma _{2458}+\widehat\Sigma _{2467}\right)\,,\\
\frak r_3^{(\alpha)} & \eql \widehat A\,{}^1{}_2+\widehat A\,{} ^3{}_4+\widehat A\,{} ^5{}_6+\widehat A\,{} ^7{}_8\,,
\end{split}
\end{equation}
\begin{equation}\label{}
\begin{split}
\frak r_1^{(\beta)} & \eql {1\over \sqrt{2}}\left(-\widehat\Sigma _{1357}+\widehat\Sigma _{1368}-\widehat\Sigma _{1458}-\widehat\Sigma _{1467}-\widehat\Sigma _{2358}-\widehat\Sigma _{2367}+\widehat\Sigma _{2457}-\widehat\Sigma _{2468}\right)\,,\\
\frak r_2^{(\beta)} & \eql {1\over \sqrt{2}}\left(\widehat\Sigma _{1358}+\widehat\Sigma _{1367}-\widehat\Sigma _{1457}+\widehat\Sigma _{1468}-\widehat\Sigma _{2357}+\widehat\Sigma _{2368}-\widehat\Sigma _{2458}-\widehat\Sigma _{2467}\right)\,,\\
\frak r_3^{(\beta)} & \eql\widehat A\,{} ^1{}_2+\widehat A\,{} ^3{}_4-\widehat A\,{} ^5{}_6-\widehat A\,{} ^7{}_8\,,
\end{split}
\end{equation}
\begin{equation}\label{}
\begin{split}
\frak r_1^{(\gamma)} & \eql {1\over \sqrt{2}}\left(-\widehat\Sigma _{1357}-\widehat\Sigma _{1368}+\widehat\Sigma _{1458}-\widehat\Sigma _{1467}-\widehat\Sigma _{2358}+\widehat\Sigma _{2367}-\widehat\Sigma _{2457}-\widehat\Sigma _{2468}\right)\,,\\
\frak r_2^{(\gamma)} & \eql {1\over \sqrt{2}}\left(\widehat\Sigma _{1358}-\widehat\Sigma _{1367}+\widehat\Sigma _{1457}+\widehat\Sigma _{1468}-\widehat\Sigma _{2357}-\widehat\Sigma _{2368}+\widehat\Sigma _{2458}-\widehat\Sigma _{2467}\right)\,,\\
\frak r_3^{(\gamma)}&\eql\widehat A\,{} ^1{}_2-\widehat A\,{} ^3{}_4+\widehat A\,{} ^5{}_6-\widehat A\,{} ^7{}_8
\end{split}
\end{equation}
\begin{equation}\label{}
\begin{split}
\frak r_1^{(\delta)} & \eql {1\over \sqrt{2}}\left(-\widehat\Sigma _{1357}-\widehat\Sigma _{1368}-\widehat\Sigma _{1458}+\widehat\Sigma _{1467}+\widehat\Sigma _{2358}-\widehat\Sigma _{2367}-\widehat\Sigma _{2457}-\widehat\Sigma _{2468}\right)\,,\\
\frak r_2^{(\delta)} & \eql {1\over \sqrt{2}}\left(\widehat\Sigma _{1358}-\widehat\Sigma _{1367}-\widehat\Sigma _{1457}-\widehat\Sigma _{1468}+\widehat\Sigma _{2357}+\widehat\Sigma _{2368}+\widehat\Sigma _{2458}-\widehat\Sigma _{2467}\right)\,,\\
\frak r_3^{(\delta)}& \eql-\widehat A\,{} ^1{}_2+\widehat A\,{} ^3{}_4+\widehat A\,{} ^5{}_6-\widehat A\,{} ^7{}_8\,,
\end{split}
\end{equation}
where $\widehat A\,{}^I{}_J=\widehat\Lambda{}^I{}_J-\widehat\Lambda{}^J{}_I$. 
They satisfy, 
\begin{equation}\label{}
[\frak r_i,\frak r_j]+4\epsilon_{ijk}\frak r_k\eql 0\,,\qquad \Tr \frak r_i\frak r_j\eql -48\,\delta_{ij}\,,
\end{equation}
within each $\su(2)$ subalgebra.

The group elements of these four commuting $\rm SU(2)$'s inside $\rm E_{6(6)}$ are now parametrized by the Euler angles $\alpha_1\,,\ldots,\delta_3$ defined by
\begin{equation}\label{}
g(\varphi_1,\varphi_2,\varphi_3)\eql \exp ({ \varphi_1\frak r_3^{(\varphi)} }) 
\cdot\exp({ \varphi_2\frak r_1^{(\varphi)} })\cdot\exp( { \varphi_3\frak r_3^{(\varphi)} })\,,\qquad \varphi=\alpha,~\beta,~\gamma,~\delta\,.
\end{equation}
By simultanously diagonalizing the four Casimir operators, one can bring the group element of the compact subgroup into a block diagonal form corresponding to the branching
\begin{equation}\label{branchsu2}
\begin{split}
{\bf 27} ~\longrightarrow ~ & 3\times {\bf (1,1,1,1)}+\\
& {\bf (2,2,1,1)+(2,1,2,1)+(1,2,2,1)+(2,1,1,2)+(1,2,1,2)+(1,1,2,2)}\,,
\end{split}
\end{equation}
where the $4\times 4$ blocks of each $\rm SU(2)$ are of the form
\begin{equation}\label{thesis}
\left(\begin{matrix}
s_1& s_2 & \sqrt 2 \,s_3 & \sqrt 2\,s_4 \\
-s_2 & s_1 & -\sqrt 2 \,s_4 & \sqrt 2\,s_3\\
-s_3/\sqrt 2 & s_4/\sqrt 2 & s_1 & -s_2 \\
-s_4/\sqrt 2 & \sqrt 3/\sqrt 2 & s_2 & s_1
\end{matrix}\right)\,,
\end{equation}
 modulo a permutation of signs between some terms that make the two $\rm SU(2)$'s in each $(\bfs 2,\bfs 2)$ block commute. 
The $s_i$'s above are  
\begin{equation}\label{}
\begin{split}
s_1 & \eql \cos 2\varphi_2\cos 2(\varphi_1+\varphi_3)\,,\qquad s_2\eql \cos 2\varphi_2\sin 2(\varphi_1+\varphi_3)\,,\\
s_3 & \eql \sin 2\varphi_2\cos 2(\varphi_1-\varphi_3)\,,\qquad s_4\eql \sin 2\varphi_2\sin 2(\varphi_1-\varphi_3)\,,\\
\end{split}
\end{equation}
for each of the angles $\varphi=\alpha,\beta,\gamma,\delta$. Note that
\begin{equation}\label{quatnr}
s_1^2+s_2^2+s_3^2+s_4^2\eql 1\,,
\end{equation}
so that each block is simply a unit quaternion. Then the $24\times 24$ block of the $\rm E_{6(6)}$ matrix corresponding to the second line in \eqref{branchsu2} is a diagonal matrix parametrized by 4 commuting quaternions, $q_\alpha,\ldots,q_\delta$.

Next we choose 4 commuting noncompact generators, cf. \eqref{trgenerators},
\begin{equation}\label{ggensKP}
\begin{split}
\frak g_1 & \eql {1\over \sqrt 2}\left(\widehat\Sigma _{1357}-\widehat\Sigma _{2468}\right)\,,\qquad 
\frak g_2   \eql {1\over \sqrt 2}\left(\widehat\Sigma _{1467}-\widehat\Sigma _{2358}\right)\,,\\
\frak g_3 & \eql {1\over \sqrt 2}\left(\widehat\Sigma _{2367}-\widehat\Sigma _{1458}\right)\,,\qquad 
\frak g_4   \eql {1\over \sqrt 2}\left(\widehat\Sigma _{2457}-\widehat\Sigma _{1368}\right)\,.\\
\end{split}
\end{equation}
Then the scalar 27-bein 
\begin{equation}\label{Vso44}
\cals V_{\SO(4,4)}(\alpha,\beta,\gamma,\delta;\rho)\eql g(\alpha)\ldots g(\delta)\exp\big({\sum_i\rho_i \frak g_i}\big)\,g(\delta)^{-1}\ldots g(\alpha)^{-1}\in {\rm E_{6(6)}}\,,
\end{equation}
parametrizes the coset $\cals M_{\SO(4,4)}$. 
The matrix $\cals V_{\SO(4,4)}$ is somewhat sparse with 195 out of $27^2=729$ nonzero entries. In the following, it will be useful to work  with the corresponding matrix obtained by replacing  the nonvanishing entries in $\cals V_{\SO(4,4)}$  with symbolic entries, say $m_{ij}$.  

Adding the $\cals M_{\rm O(1,1)}$ factor does not change much. We choose generators, cf. \eqref{galbe},
\begin{equation}\label{o11generators}
\begin{split}
\tilde{\frak g}_1 & \eql {3\over 2}\left(\widehat\Lambda{} ^1{}_1+\widehat\Lambda{} ^2{}_2-\widehat\Lambda{} ^3{}_3-\widehat\Lambda{} ^4{}_4\right)\,,\\
\tilde {\frak g}_2 & \eql {5\over 3}\left(\widehat\Lambda{} ^1{}_1+\widehat\Lambda{} ^2{}_2+\widehat\Lambda{} ^3{}_3+\widehat\Lambda{} ^4{}_4-2 \widehat\Lambda{} ^5{}_5-2 \widehat\Lambda{} ^6{}_6\right)\,,
\end{split}
\end{equation}
with the corresponding group element
\begin{equation}\label{}
\cals V_{\rm O(1,1)^2}(\xi_1,\xi_2)\eql \exp({\xi_1\, \tilde{\frak g}_1+\xi_2\, \tilde{\frak g}_2})\,.
\end{equation}
This matrix is diagonal and simply ``decorates'' the $m_{ij}$'s  in \eqref{Vso44} by exponential factors. Finally, the full scalar 27-bein is
\begin{equation}\label{}
\cals V(\xi;\alpha,\beta,\gamma,\delta;\rho)\eql \cals V_{\rm O(1,1)^2}(\xi)\cdot \cals V_\text{SO(4,4)}(\alpha,\beta,\gamma,\delta;\rho)\,.
\end{equation}

\subsection{Computation of the potential}

Using symbolic representation of $\cals V$, the potential is a sum of 2784 terms quartic in $m_{ij}$'s, which fall into 6 different groups depending on the $\rm O(1,1)^2$ prefactors,
\begin{equation}\label{prefxi}
e^{-3 \xi _1-\frac{10 }{3}\xi _2},\qquad e^{3 \xi _1-\frac{10 }{3}\xi _2},\qquad e^{-\frac{40 }{3}\xi _2},\qquad e^{\frac{20 }{3}\xi
   _2},\qquad e^{-6 \xi _1+\frac{20 }{3}\xi _2},\qquad e^{6 \xi _1+\frac{20}{3} \xi _2}\,.
\end{equation}
After substituting for  $m_{ij}$'s, we find the prefactors  in \eqref{prefxi} are multiplied by 48, 48, 18, 48, 18, 18 different quartic products of $\cosh\rho_{ij}$ and $\sinh\rho_{ij}$, $\rho_{ij}=\rho_i-\rho_j$, respectively, for the total of 198 terms. In turn, each of those terms is multiplied by a homogenous polynomial of order 16 in 16 different $s_i$'s \eqref{thesis} for the 12 Euler angles. A typical number of terms in those trigonometric polynomials is on the order of 40,000. That number is drastically reduced upon repeated use of \eqref{quatnr}, usually to less than a 100. Finally, the substitution  of explicit $s_i$'s in terms of the angles futher collapses each group to a relatively small number of terms.

In the last stage, all dependence of the potential on the four angles $\alpha_1$, $\beta_1$, $\gamma_1$ and $\delta_1$ disappears and one is left with the potential that depends on 8 Euler angles and 6 noncompact fields. This is a nice consistency check for this long calculation. The $\ZZ_2\times \ZZ_2$ truncation  preserves  $\rm \U(1)^4\subset \SO(6)\times \SL(2,\RR)$, generated by $\frak r_3^{(\varphi)}$'s, which is a symmetry of the potential. Hence the latter should   be a function of $18-4=14$ independent scalar fields, as indeed it is. 

Even a simplified expression for the potential is too long to be written down in a reasonable amount of space here.  Instead, it is made available as a Mathematica input file, see Section~\ref{App:PotFile}.

\subsection{The critical points}
\label{App:so44points}

\begin{table}[t]
\begin{center}\tt
\begin{tabular}{@{\extracolsep{10 pt}} cccccc}
\toprule
 T0750000 & 
 T0780031   & T0839947   & T0843750  & T0870297 & 
 T0878939  \\
  T0887636  & T0892913   & T0964525  & 
 T0982778   & T1001482 & T1054687 \\ 
 T1125000  & 
T1304606   & T1417411  & T1501862 & T1547778 \\
\noalign{\smallskip}
\bottomrule
\end{tabular}
\caption{\label{so44points}
The critical points in the $\ZZ_2\times \ZZ_2$-invariant sector.}
\end{center}
\end{table}

We have found 17 critical points of the scalar potential in this $\ZZ_2\times\ZZ_2$-invariant sector using the {$\tt FindRoot[~\cdot~]$} routine in Mathematica starting at random locations on the scalar manifold. Those points are listed in Table~\ref{so44points}. As expected, they include all critical points found in the 10-scalar model in Section~\ref{sec:truncation}, with only two additional ones, $\tt T1054687$ and 
$\tt T1547778$, whose positions in the polar parametrization used here are given in Table~\ref{twopoints}.\footnote{Note that
\begin{equation*}
-{ 3^{5/3}\cdot 5\over  2^{13/3}}\eql -1.5477783979193562580662234151917585735219771770242937517061887\dots\,,
\end{equation*}
which agrees with the value of the potential for $\tt T1547778$ to the numerical accuracy we tested it.}

A major inconvenience when working with the polar-type coordinates, as compared to the ones used in Sections~\ref{sec:truncation} and \ref{sec:TFAdS}, is the presence of coordinate singularities in the parametrization of the scalar coset. As a result the search routine yields a large fraction of ``fake critical points''.  Those are then eliminated by an explicit check of criticality, that is  by evaluating the potential to the first order in $\epsilon$  on the scalar vielbein
\begin{equation}\label{varT}
(1+\epsilon \sum_A \psi_A T_A)\,\cals V_*\,,
\end{equation}
where $\cals V_*$ is the presumed critical point. The sum in \eqref{varT} runs over all 78 generators of $\frak e_{6(6)}$ and to eliminate a point it is sufficient to verify that \eqref{varT} does not vanish for some random values of the parameters $\psi_A$.

The numerical search for ctitical points in this sector appears to be quite efficient, so we believe that there should be no missing critical points from our search. 
It is then quite  remarkable that the 17 points found here constitute more than 50\% of all critical points found by the TensorFlow search in Section~\ref{sec:TFAdS}.  

\begin{table}[t]
\begin{center}
\resizebox{0.95\textwidth}{!}{%
\begin{tabular}{@{\extracolsep{0 pt}} c cc cccc cc cc cc cc }
\toprule
Point & $\xi_1$ & $\xi_2$ & $\rho_1$ &  $\rho_2$ & $\rho_3$ & $\rho_4$ & $\alpha_{2/3}$  & 
$\beta_{2/3}$ & $\gamma_{2/3}$ & $\delta_{2/3}$ \\
\noalign{\smallskip}
\midrule
\noalign{\smallskip}
{\tt T1054687} & 0 &  0 &  0.59672 & $ -0.00571$ &  0.60944&  0.59486&  0.05047&  0.38645&  0.52874 & 
0.28433 \\
& & & &&&& 0.46797 &  1.17929 &  1.64474 &  0.85757\\[6 pt]
{\tt T1547778} & 
0.14931 &  0.04479 &  0.34047 &  1.36783 &  0.56949 &  0.58893 &  0.58938 &   
1.5804 &  1.05899 &  1.38975\\
&&&&&&& 0.23497&  0.41012 &  1.56041 &  1.36012\\
\noalign{\smallskip}
\bottomrule
\end{tabular}
}
\caption{\label{twopoints}
Positions of {\tt T1054687} and {\tt T1547778} in the polar coordinates.}
\end{center}
\end{table}

\subsection{Ancillary files}
\label{App:PotFile}

A text file with a Mathematica input for the full potential in the $\ZZ_2\times \ZZ_2$-invariant sector in this section is available for download as an ancillary file with this \texttt{arXiv} submission.
The potential depends on  14 scalar fields, which are denoted by the same symbols as in 
the text above. The locations of the critical points can be found in a Mathematica input file which is available for download as an ancillary file with this \texttt{arXiv} submission.



\section{Critical points and mass spectra}
\label{appendixB}

In this appendix, we list numerical data obtained from TensorFlow on
the locations, gravitino and scalar mass spectra, cosmological
constant, as well as residual gauge symmetry and
supersymmetry. Gravitino masses are normalized relative to the AdS
radius such that for every unbroken supersymmetry, there is
a~$m^2/m_0^2[\psi]=1$ gravitino, and the BF bound
is~$m^2/m_0^2[\phi]\ge-4$. The
(symmetric-traceless)~$\Lambda^{I}{}_{J}$ parameters have been
diagonalized. The diagonal~$\Lambda^{I}{}_{I}$ entries listed sum to
zero as expected, hiding a linear constraint on the numerical data.

This list of solutions, produced by starting numerical optimization
from~$10^5$ random points, is likely to be mostly complete. Notably,
an independent second deep scan that used a modified `loss function'
to guide the search towards supersymmetric solutions did not find any
supersymmetric critical points beyond the two already known ones.

In the list of solutions we give the location on the scalar manifold in terms of the generators in Appendix~\ref{appConv}. The $\mathfrak{e}_{6(6)}$ element is constructed as a linear combination of the generators $\widehat{\Lambda}^\alpha{}_\beta$, $\widehat{\Lambda}^I{}_J$, and $\widehat{\Sigma}_{IJK\alpha}$. The coefficient of $\widehat{\Lambda}^\alpha{}_\beta$ is set to zero for all solutions as explained in Section~\ref{sec:TFAdS}. The coefficients of $\widehat{\Lambda}^I{}_J$ are denoted by ${\Lambda}^I{}_J$, and the coefficients multiplying $\widehat{\Sigma}_{IJK7}$ are $\pm\sqrt{2} \Sigma_{\pm (IJK;1+\cdots;2)}$. Only nonzero coefficients of the $\widehat{\Sigma}_{IJK\alpha}$-generators are displayed. This accounts for all non-compact generators. The group element is obtained by exponentiating the linear combination just described.

\begin{flalign}{\bf T0750000}: V/g^2 = -0.75000000,\;\mathcal{N}=8,\;\mathfrak{so}(6)\to\mathfrak{su}(4)&&\end{flalign}\\[-8.5ex]
\begin{flalign*}\begin{autobreak}m^2/m_0^2[\psi]: 
1.000_{\times 8}\end{autobreak}&&\end{flalign*}\\[-8.5ex]
\begin{flalign*}\begin{autobreak}m^2/m_0^2[\phi]: 
-4.000_{\times 20},
-3.000_{\times 20},
0.000_{\times 2}\end{autobreak}&&\end{flalign*}\\[-8.5ex]
\begin{flalign*}\begin{autobreak}\kern0.0001em
\Lambda^1{}_1=\Lambda^{2}{}_{2}=\Lambda^{3}{}_{3}=\Lambda^{4}{}_{4}=\Lambda^{5}{}_{5}=\Lambda^{6}{}_{6}=0\end{autobreak}&&\end{flalign*}\\[-8.5ex]
\\[3ex]

\hrule
\begin{flalign}{\bf T0780031}: V/g^2 = -0.78003143,\;\mathfrak{so}(6)\to\mathfrak{so}(5)&&\end{flalign}\\[-8.5ex]
\begin{flalign*}\begin{autobreak}m^2/m_0^2[\psi]: 
1.185_{\times 8}\end{autobreak}&&\end{flalign*}\\[-8.5ex]
\begin{flalign*}\begin{autobreak}m^2/m_0^2[\phi]: 
-5.333_{\times 14},
-2.000_{\times 20},
0.000_{\times 7},
8.000\end{autobreak}&&\end{flalign*}\\[-8.5ex]
\begin{flalign*}\begin{autobreak}\kern0.0001em
\Lambda^1{}_1=\Lambda^{2}{}_{2}=\Lambda^{3}{}_{3}=\Lambda^{4}{}_{4}=\Lambda^{5}{}_{5}\approx-0.09155,
\Lambda^{6}{}_{6}\approx0.45776\end{autobreak}&&\end{flalign*}\\[-8.5ex]
\\[3ex]

\hrule
\begin{flalign}{\bf T0839947}: V/g^2 = -0.83994737,\;\mathcal{N}=2,\;\mathfrak{so}(6)\to\mathfrak{su}(2) + \mathfrak{u}(1)&&\end{flalign}\\[-8.5ex]
\begin{flalign*}\begin{autobreak}m^2/m_0^2[\psi]: 
1.000_{\times 2},
1.361_{\times 4},
1.778_{\times 2}\end{autobreak}&&\end{flalign*}\\[-8.5ex]
\begin{flalign*}\begin{autobreak}m^2/m_0^2[\phi]: 
-4.000_{\times 3},
-3.750_{\times 12},
-3.437_{\times 4},
-3.000_{\times 2},
-2.438_{\times 4},
-1.292,
0.000_{\times 13},
3.000_{\times 2},
9.292\end{autobreak}&&\end{flalign*}\\[-8.5ex]
\begin{flalign*}\begin{autobreak}\kern0.0001em
\Lambda^1{}_1=\Lambda^{2}{}_{2}=\Lambda^{3}{}_{3}=\Lambda^{4}{}_{4}\approx-0.11552,
\Lambda^{5}{}_{5}=\Lambda^{6}{}_{6}\approx0.23105\end{autobreak}&&\end{flalign*}\\[-8.5ex]
\begin{flalign*}\begin{autobreak}\kern0.0001em
\Sigma_{+125;1+346;2}\approx-0.27465,
\Sigma_{+126;2+345;1}=\Sigma_{+135;2+246;1}=\Sigma_{+136;1+245;2}\approx0.27465\end{autobreak}&&\end{flalign*}\\[-8.5ex]
\\[3ex]

\hrule
\begin{flalign}{\bf T0843750}: V/g^2 = -0.84375000,\;\mathfrak{so}(6)\to\mathfrak{su}(3)&&\end{flalign}\\[-8.5ex]
\begin{flalign*}\begin{autobreak}m^2/m_0^2[\psi]: 
1.210_{\times 6},
2.000_{\times 2}\end{autobreak}&&\end{flalign*}\\[-8.5ex]
\begin{flalign*}\begin{autobreak}m^2/m_0^2[\phi]: 
-4.444_{\times 12},
-1.778_{\times 12},
0.000_{\times 17},
8.000\end{autobreak}&&\end{flalign*}\\[-8.5ex]
\begin{flalign*}\begin{autobreak}\kern0.0001em
\Lambda^1{}_1=\Lambda^{2}{}_{2}=\Lambda^{3}{}_{3}=\Lambda^{4}{}_{4}=\Lambda^{5}{}_{5}=\Lambda^{6}{}_{6}\approx0.00000\end{autobreak}&&\end{flalign*}\\[-8.5ex]
\begin{flalign*}\begin{autobreak}\kern0.0001em
\Sigma_{+123;1+456;2}\approx0.29343,
\Sigma_{+123;2-456;1}\approx0.02565,
\Sigma_{+124;1-356;2}\approx0.00727,
\Sigma_{+124;2+356;1}\approx0.02437,
\Sigma_{+125;1+346;2}\approx0.01975,
\Sigma_{+125;2-346;1}\approx-0.30327,
\Sigma_{+126;1-345;2}\approx-0.08422,
\Sigma_{+126;2+345;1}\approx0.02034,
\Sigma_{+134;1+256;2}\approx0.01948,
\Sigma_{+134;2-256;1}\approx-0.31733,
\Sigma_{+135;1-246;2}\approx0.00577,
\Sigma_{+135;2+246;1}\approx0.00693,
\Sigma_{+136;1+245;2}\approx0.12256,
\Sigma_{+136;2-245;1}\approx-0.01130,
\Sigma_{+145;1+236;2}\approx0.32741,
\Sigma_{+145;2-236;1}\approx0.01353,
\Sigma_{+146;1-235;2}\approx-0.01950,
\Sigma_{+146;2+235;1}\approx-0.08083,
\Sigma_{+156;1+234;2}\approx-0.01220,
\Sigma_{+156;2-234;1}\approx-0.12477\end{autobreak}&&\end{flalign*}\\[-8.5ex]
\\[3ex]

\hrule
\begin{flalign}{\bf T0870297}: V/g^2 = -0.87029791,\;\mathfrak{so}(6)\to\mathfrak{su}(2) + \mathfrak{u}(1)&&\end{flalign}\\[-8.5ex]
\begin{flalign*}\begin{autobreak}m^2/m_0^2[\psi]: 
1.440_{\times 4},
1.600_{\times 4}\end{autobreak}&&\end{flalign*}\\[-8.5ex]
\begin{flalign*}\begin{autobreak}m^2/m_0^2[\phi]: 
-5.440_{\times 6},
-4.000_{\times 4},
-2.560_{\times 8},
-2.400_{\times 6},
0.000_{\times 13},
3.360_{\times 2},
9.600,
10.400_{\times 2}\end{autobreak}&&\end{flalign*}\\[-8.5ex]
\begin{flalign*}\begin{autobreak}\kern0.0001em
\Lambda^1{}_1=\Lambda^{2}{}_{2}=\Lambda^{3}{}_{3}=\Lambda^{4}{}_{4}\approx-0.19188,
\Lambda^{5}{}_{5}=\Lambda^{6}{}_{6}\approx0.38376\end{autobreak}&&\end{flalign*}\\[-8.5ex]
\begin{flalign*}\begin{autobreak}\kern0.0001em
\Sigma_{+135;2+246;1}=\Sigma_{+136;1+245;2}\approx-0.35635\end{autobreak}&&\end{flalign*}\\[-8.5ex]
\\[3ex]

\hrule
\begin{flalign}{\bf T0878939}: V/g^2 = -0.87893974,\;\mathfrak{so}(6)\to\mathfrak{u}(1)&&\end{flalign}\\[-8.5ex]
\begin{flalign*}\begin{autobreak}m^2/m_0^2[\psi]: 
1.358_{\times 4},
1.802_{\times 4}\end{autobreak}&&\end{flalign*}\\[-8.5ex]
\begin{flalign*}\begin{autobreak}m^2/m_0^2[\phi]: 
-5.827,
-5.221_{\times 4},
-4.990,
-4.937_{\times 2},
-4.780_{\times 2},
-4.475,
-2.522_{\times 2},
-2.288_{\times 4},
-1.532_{\times 4},
0.000_{\times 16},
0.820,
4.437,
9.803,
12.622_{\times 2}\end{autobreak}&&\end{flalign*}\\[-8.5ex]
\begin{flalign*}\begin{autobreak}\kern0.0001em
\Lambda^1{}_1=\Lambda^{2}{}_{2}\approx-0.21480,
\Lambda^{3}{}_{3}=\Lambda^{4}{}_{4}\approx-0.20031,
\Lambda^{5}{}_{5}=\Lambda^{6}{}_{6}\approx0.41511\end{autobreak}&&\end{flalign*}\\[-8.5ex]
\begin{flalign*}\begin{autobreak}\kern0.0001em
\Sigma_{+136;2-245;1}=\Sigma_{+146;1-235;2}\approx0.34860\end{autobreak}&&\end{flalign*}\\[-8.5ex]
\\[3ex]

\hrule
\begin{flalign}{\bf T0887636}: V/g^2 = -0.88763615,\;\mathfrak{so}(6)\to\mathfrak{u}(1)&&\end{flalign}\\[-8.5ex]
\begin{flalign*}\begin{autobreak}m^2/m_0^2[\psi]: 
1.483_{\times 2},
1.604_{\times 4},
1.838_{\times 2}\end{autobreak}&&\end{flalign*}\\[-8.5ex]
\begin{flalign*}\begin{autobreak}m^2/m_0^2[\phi]: 
-5.907_{\times 2},
-5.489_{\times 4},
-5.208,
-4.731,
-4.581_{\times 2},
-4.054_{\times 2},
-2.293_{\times 4},
-1.182_{\times 2},
-0.253_{\times 2},
0.000_{\times 16},
2.070_{\times 2},
3.344,
9.651,
14.863_{\times 2}\end{autobreak}&&\end{flalign*}\\[-8.5ex]
\begin{flalign*}\begin{autobreak}\kern0.0001em
\Lambda^1{}_1=\Lambda^{2}{}_{2}=\Lambda^{3}{}_{3}=\Lambda^{4}{}_{4}\approx-0.22251,
\Lambda^{5}{}_{5}=\Lambda^{6}{}_{6}\approx0.44502\end{autobreak}&&\end{flalign*}\\[-8.5ex]
\begin{flalign*}\begin{autobreak}\kern0.0001em
\Sigma_{+125;1+346;2}\approx0.16666,
\Sigma_{+126;2+345;1}\approx0.03828,
\Sigma_{+135;2+246;1}\approx-0.14968,
\Sigma_{+136;1+245;2}\approx0.14968,
\Sigma_{+145;1+236;2}\approx-0.15351,
\Sigma_{+146;2+235;1}\approx0.37172\end{autobreak}&&\end{flalign*}\\[-8.5ex]
\\[3ex]

\hrule
\begin{flalign}{\bf T0892913}: V/g^2 = -0.89291309,\;\mathfrak{so}(6)\to\mathfrak{u}(1) + \mathfrak{u}(1)&&\end{flalign}\\[-8.5ex]
\begin{flalign*}\begin{autobreak}m^2/m_0^2[\psi]: 
1.667_{\times 8}\end{autobreak}&&\end{flalign*}\\[-8.5ex]
\begin{flalign*}\begin{autobreak}m^2/m_0^2[\phi]: 
-6.000_{\times 4},
-5.572,
-5.000_{\times 4},
-4.520,
-3.500_{\times 4},
-0.833_{\times 4},
0.000_{\times 16},
1.667_{\times 4},
2.667,
9.572,
16.000,
16.520\end{autobreak}&&\end{flalign*}\\[-8.5ex]
\begin{flalign*}\begin{autobreak}\kern0.0001em
\Lambda^1{}_1=\Lambda^{2}{}_{2}=\Lambda^{3}{}_{3}=\Lambda^{4}{}_{4}\approx-0.23105,
\Lambda^{5}{}_{5}=\Lambda^{6}{}_{6}\approx0.46210\end{autobreak}&&\end{flalign*}\\[-8.5ex]
\begin{flalign*}\begin{autobreak}\kern0.0001em
\Sigma_{+125;1+346;2}\approx-0.00001,
\Sigma_{+125;2-346;1}\approx0.00001,
\Sigma_{+126;1-345;2}\approx-0.00198,
\Sigma_{+126;2+345;1}\approx-0.00217,
\Sigma_{+135;1-246;2}\approx0.35483,
\Sigma_{+135;2+246;1}\approx-0.32504\end{autobreak}&&\end{flalign*}\\[-8.5ex]
\\[3ex]

\hrule
\begin{flalign}{\bf T0963952}: V/g^2 = -0.96395224,\;\mathfrak{so}(6)\to\emptyset&&\end{flalign}\\[-8.5ex]
\begin{flalign*}\begin{autobreak}m^2/m_0^2[\psi]: 
1.624_{\times 2},
1.753_{\times 2},
1.924_{\times 2},
2.063_{\times 2}\end{autobreak}&&\end{flalign*}\\[-8.5ex]
\begin{flalign*}\begin{autobreak}m^2/m_0^2[\phi]: 
-5.874,
-5.026_{\times 2},
-4.449_{\times 2},
-4.076_{\times 2},
-4.017_{\times 2},
-4.010_{\times 2},
-3.485_{\times 2},
-2.623,
-0.499,
-0.156_{\times 2},
0.000_{\times 17},
6.042_{\times 2},
7.106_{\times 2},
9.730,
11.012_{\times 2},
12.757\end{autobreak}&&\end{flalign*}\\[-8.5ex]
\begin{flalign*}\begin{autobreak}\kern0.0001em
\Lambda^1{}_1\approx-0.32137,
\Lambda^{2}{}_{2}=\Lambda^{3}{}_{3}\approx-0.24369,
\Lambda^{4}{}_{4}=\Lambda^{5}{}_{5}\approx0.24012,
\Lambda^{6}{}_{6}\approx0.32852\end{autobreak}&&\end{flalign*}\\[-8.5ex]
\begin{flalign*}\begin{autobreak}\kern0.0001em
\Sigma_{+126;2+345;1}=\Sigma_{+136;1+245;2}\approx0.17139,
\Sigma_{+146;1-235;2}=\Sigma_{+156;2-234;1}\approx0.52343\end{autobreak}&&\end{flalign*}\\[-8.5ex]
\\[3ex]

\hrule
\begin{flalign}{\bf T0964097}: V/g^2 = -0.96409727,\;\mathfrak{so}(6)\to\emptyset&&\end{flalign}\\[-8.5ex]
\begin{flalign*}\begin{autobreak}m^2/m_0^2[\psi]: 
1.701_{\times 4},
1.985_{\times 4}\end{autobreak}&&\end{flalign*}\\[-8.5ex]
\begin{flalign*}\begin{autobreak}m^2/m_0^2[\phi]: 
-5.922,
-5.120,
-5.100,
-4.711,
-4.163,
-4.146,
-4.097,
-4.056,
-4.040,
-3.996,
-3.943,
-3.455,
-3.450,
-2.553,
-0.456,
0.000_{\times 18},
0.137,
5.735,
5.821,
7.182,
7.379,
9.682,
10.417,
11.561,
13.019\end{autobreak}&&\end{flalign*}\\[-8.5ex]
\begin{flalign*}\begin{autobreak}\kern0.0001em
\Lambda^1{}_1\approx-0.32305,
\Lambda^{2}{}_{2}\approx-0.24370,
\Lambda^{3}{}_{3}\approx-0.23916,
\Lambda^{4}{}_{4}\approx0.22698,
\Lambda^{5}{}_{5}\approx0.24757,
\Lambda^{6}{}_{6}\approx0.33137\end{autobreak}&&\end{flalign*}\\[-8.5ex]
\begin{flalign*}\begin{autobreak}\kern0.0001em
\Sigma_{+136;2-245;1}\approx0.21467,
\Sigma_{+146;2+235;1}\approx-0.51776,
\Sigma_{+156;1+234;2}\approx0.54357\end{autobreak}&&\end{flalign*}\\[-8.5ex]
\\[3ex]

\hrule
\begin{flalign}{\bf T0964525}: V/g^2 = -0.96452592,\;\mathfrak{so}(6)\to\mathfrak{u}(1)&&\end{flalign}\\[-8.5ex]
\begin{flalign*}\begin{autobreak}m^2/m_0^2[\psi]: 
1.798_{\times 4},
1.899_{\times 4}\end{autobreak}&&\end{flalign*}\\[-8.5ex]
\begin{flalign*}\begin{autobreak}m^2/m_0^2[\phi]: 
-5.981,
-5.350_{\times 2},
-4.441_{\times 2},
-4.119_{\times 4},
-3.992_{\times 2},
-3.453_{\times 2},
-2.393,
0.000_{\times 16},
0.275_{\times 4},
5.108_{\times 2},
7.699_{\times 2},
9.546,
11.108_{\times 2},
13.615\end{autobreak}&&\end{flalign*}\\[-8.5ex]
\begin{flalign*}\begin{autobreak}\kern0.0001em
\Lambda^1{}_1\approx-0.32814,
\Lambda^{2}{}_{2}=\Lambda^{3}{}_{3}\approx-0.23391,
\Lambda^{4}{}_{4}=\Lambda^{5}{}_{5}\approx0.22848,
\Lambda^{6}{}_{6}\approx0.33900\end{autobreak}&&\end{flalign*}\\[-8.5ex]
\begin{flalign*}\begin{autobreak}\kern0.0001em
\Sigma_{+146;2+235;1}=\Sigma_{+156;1+234;2}\approx-0.55645\end{autobreak}&&\end{flalign*}\\[-8.5ex]
\\[3ex]

\hrule
\begin{flalign}{\bf T0982778}: V/g^2 = -0.98277802,\;\mathfrak{so}(6)\to\mathfrak{u}(1)&&\end{flalign}\\[-8.5ex]
\begin{flalign*}\begin{autobreak}m^2/m_0^2[\psi]: 
1.630_{\times 4},
2.222_{\times 4}\end{autobreak}&&\end{flalign*}\\[-8.5ex]
\begin{flalign*}\begin{autobreak}m^2/m_0^2[\phi]: 
-6.371,
-5.431,
-5.333_{\times 2},
-4.114_{\times 2},
-4.000_{\times 2},
-3.277,
-3.033_{\times 2},
-2.846,
-1.903,
-1.127_{\times 2},
0.000_{\times 17},
2.194_{\times 2},
6.611,
7.033_{\times 2},
9.333,
9.799,
11.714_{\times 2},
14.085\end{autobreak}&&\end{flalign*}\\[-8.5ex]
\begin{flalign*}\begin{autobreak}\kern0.0001em
\Lambda^1{}_1=\Lambda^{2}{}_{2}\approx-0.32349,
\Lambda^{3}{}_{3}\approx-0.19823,
\Lambda^{4}{}_{4}=\Lambda^{5}{}_{5}\approx0.25591,
\Lambda^{6}{}_{6}\approx0.33339\end{autobreak}&&\end{flalign*}\\[-8.5ex]
\begin{flalign*}\begin{autobreak}\kern0.0001em
\Sigma_{+135;2+246;1}=\Sigma_{+156;1+234;2}\approx0.56869\end{autobreak}&&\end{flalign*}\\[-8.5ex]
\\[3ex]

\hrule
\begin{flalign}{\bf T1001482}: V/g^2 = -1.00148265,\;\mathfrak{so}(6)\to\mathfrak{u}(1)&&\end{flalign}\\[-8.5ex]
\begin{flalign*}\begin{autobreak}m^2/m_0^2[\psi]: 
1.744_{\times 4},
2.188_{\times 4}\end{autobreak}&&\end{flalign*}\\[-8.5ex]
\begin{flalign*}\begin{autobreak}m^2/m_0^2[\phi]: 
-6.436,
-5.586_{\times 4},
-4.410_{\times 2},
-2.824,
-2.719_{\times 2},
-2.306_{\times 4},
0.000_{\times 16},
3.171_{\times 2},
5.354_{\times 2},
7.693_{\times 4},
8.788,
9.401,
11.392,
11.598\end{autobreak}&&\end{flalign*}\\[-8.5ex]
\begin{flalign*}\begin{autobreak}\kern0.0001em
\Lambda^1{}_1=\Lambda^{2}{}_{2}\approx-0.34956,
\Lambda^{3}{}_{3}=\Lambda^{4}{}_{4}\approx0.15576,
\Lambda^{5}{}_{5}=\Lambda^{6}{}_{6}\approx0.19379\end{autobreak}&&\end{flalign*}\\[-8.5ex]
\begin{flalign*}\begin{autobreak}\kern0.0001em
\Sigma_{+136;2-245;1}=\Sigma_{+146;1-235;2}\approx-0.63157\end{autobreak}&&\end{flalign*}\\[-8.5ex]
\\[3ex]

\hrule
\begin{flalign}{\bf T1054687}: V/g^2 = -1.05468750,\;\mathfrak{so}(6)\to\emptyset&&\end{flalign}\\[-8.5ex]
\begin{flalign*}\begin{autobreak}m^2/m_0^2[\psi]: 
1.733_{\times 2},
2.326_{\times 6}\end{autobreak}&&\end{flalign*}\\[-8.5ex]
\begin{flalign*}\begin{autobreak}m^2/m_0^2[\phi]: 
-6.512,
-5.251_{\times 2},
-4.161_{\times 3},
-4.062_{\times 2},
-2.925_{\times 2},
-1.560_{\times 3},
0.000_{\times 18},
4.490_{\times 2},
6.697_{\times 3},
10.600,
15.735_{\times 3},
16.638_{\times 2}\end{autobreak}&&\end{flalign*}\\[-8.5ex]
\begin{flalign*}\begin{autobreak}\kern0.0001em
\Lambda^1{}_1=\Lambda^{2}{}_{2}=\Lambda^{3}{}_{3}\approx-0.37177,
\Lambda^{4}{}_{4}=\Lambda^{5}{}_{5}=\Lambda^{6}{}_{6}\approx0.37177\end{autobreak}&&\end{flalign*}\\[-8.5ex]
\begin{flalign*}\begin{autobreak}\kern0.0001em
\Sigma_{+124;1-356;2}\approx-0.40776,
\Sigma_{+124;2+356;1}\approx-0.09457,
\Sigma_{+125;2-346;1}\approx-0.20112,
\Sigma_{+126;1-345;2}\approx-0.07976,
\Sigma_{+126;2+345;1}\approx0.00066,
\Sigma_{+134;1+256;2}\approx0.04717,
\Sigma_{+134;2-256;1}\approx0.21114,
\Sigma_{+135;1-246;2}\approx0.34833,
\Sigma_{+135;2+246;1}\approx-0.16791,
\Sigma_{+136;1+245;2}\approx-0.06930,
\Sigma_{+136;2-245;1}\approx-0.14455,
\Sigma_{+145;1+236;2}\approx-0.43585,
\Sigma_{+146;2+235;1}\approx-0.17902,
\Sigma_{+156;1+234;2}\approx-0.00305\end{autobreak}&&\end{flalign*}\\[-8.5ex]
\\[3ex]

\hrule
\begin{flalign}{\bf T1073529}: V/g^2 = -1.07352975,\;\mathfrak{so}(6)\to\emptyset&&\end{flalign}\\[-8.5ex]
\begin{flalign*}\begin{autobreak}m^2/m_0^2[\psi]: 
1.819_{\times 2},
2.011_{\times 2},
2.433_{\times 2},
2.806_{\times 2}\end{autobreak}&&\end{flalign*}\\[-8.5ex]
\begin{flalign*}\begin{autobreak}m^2/m_0^2[\phi]: 
-6.536,
-4.882_{\times 2},
-4.631_{\times 2},
-4.068_{\times 2},
-3.714_{\times 2},
-3.354,
-2.637_{\times 2},
0.000_{\times 17},
1.273_{\times 2},
5.470_{\times 2},
8.041,
8.130_{\times 2},
10.762,
14.391_{\times 2},
19.457_{\times 2},
21.617\end{autobreak}&&\end{flalign*}\\[-8.5ex]
\begin{flalign*}\begin{autobreak}\kern0.0001em
\Lambda^1{}_1=\Lambda^{2}{}_{2}\approx-0.39487,
\Lambda^{3}{}_{3}\approx-0.35452,
\Lambda^{4}{}_{4}=\Lambda^{5}{}_{5}\approx0.34969,
\Lambda^{6}{}_{6}\approx0.44490\end{autobreak}&&\end{flalign*}\\[-8.5ex]
\begin{flalign*}\begin{autobreak}\kern0.0001em
\Sigma_{+124;2+356;1}\approx0.47010,
\Sigma_{+125;1+346;2}\approx-0.47010,
\Sigma_{+134;1+256;2}=\Sigma_{+146;2+235;1}=\Sigma_{+156;1+234;2}\approx-0.25201,
\Sigma_{+135;2+246;1}\approx0.25201\end{autobreak}&&\end{flalign*}\\[-8.5ex]
\\[3ex]

\hrule
\begin{flalign}{\bf T1125000}: V/g^2 = -1.12500000,\;\mathfrak{so}(6)\to\emptyset&&\end{flalign}\\[-8.5ex]
\begin{flalign*}\begin{autobreak}m^2/m_0^2[\psi]: 
2.444_{\times 8}\end{autobreak}&&\end{flalign*}\\[-8.5ex]
\begin{flalign*}\begin{autobreak}m^2/m_0^2[\phi]: 
-7.325,
-6.206,
-4.769,
-4.748_{\times 2},
-4.427,
-3.550_{\times 2},
-3.280,
-1.276,
0.000_{\times 18},
4.806_{\times 2},
4.883_{\times 2},
5.333,
6.504,
7.861_{\times 2},
10.829,
16.719,
18.634,
22.748_{\times 2},
25.265\end{autobreak}&&\end{flalign*}\\[-8.5ex]
\begin{flalign*}\begin{autobreak}\kern0.0001em
\Lambda^1{}_1\approx-0.48597,
\Lambda^{2}{}_{2}=\Lambda^{3}{}_{3}\approx-0.39170,
\Lambda^{4}{}_{4}=\Lambda^{5}{}_{5}\approx0.39170,
\Lambda^{6}{}_{6}\approx0.48597\end{autobreak}&&\end{flalign*}\\[-8.5ex]
\begin{flalign*}\begin{autobreak}\kern0.0001em
\Sigma_{+125;1+346;2}\approx0.60475,
\Sigma_{+134;2-256;1}\approx-0.60475\end{autobreak}&&\end{flalign*}\\[-8.5ex]
\\[3ex]

\hrule
\begin{flalign}{\bf T1297247}: V/g^2 = -1.29724786,\;\mathfrak{so}(6)\to\emptyset&&\end{flalign}\\[-8.5ex]
\begin{flalign*}\begin{autobreak}m^2/m_0^2[\psi]: 
1.901_{\times 2},
2.568_{\times 2},
2.964_{\times 2},
3.758_{\times 2}\end{autobreak}&&\end{flalign*}\\[-8.5ex]
\begin{flalign*}\begin{autobreak}m^2/m_0^2[\phi]: 
-5.445,
-4.874,
-4.060,
-3.966,
-3.942,
-3.204,
-3.186,
-2.907,
-1.014,
-0.934,
0.000_{\times 18},
6.983,
7.652,
11.889,
12.146,
12.543,
13.041,
13.946,
14.490,
18.175,
18.187,
21.001,
21.574,
22.590,
22.900\end{autobreak}&&\end{flalign*}\\[-8.5ex]
\begin{flalign*}\begin{autobreak}\kern0.0001em
\Lambda^1{}_1=\Lambda^{2}{}_{2}\approx-0.59541,
\Lambda^{3}{}_{3}=\Lambda^{4}{}_{4}\approx0.23362,
\Lambda^{5}{}_{5}=\Lambda^{6}{}_{6}\approx0.36178\end{autobreak}&&\end{flalign*}\\[-8.5ex]
\begin{flalign*}\begin{autobreak}\kern0.0001em
\Sigma_{+123;1+456;2}\approx0.48716,
\Sigma_{+124;2+356;1}\approx-0.48716,
\Sigma_{+125;1+346;2}=\Sigma_{+126;2+345;1}\approx0.04462,
\Sigma_{+134;1+256;2}=\Sigma_{+156;1+234;2}\approx-0.37352,
\Sigma_{+135;2+246;1}\approx0.28567,
\Sigma_{+136;1+245;2}\approx0.33709,
\Sigma_{+145;1+236;2}\approx-0.47047,
\Sigma_{+146;2+235;1}\approx0.03521\end{autobreak}&&\end{flalign*}\\[-8.5ex]
\\[3ex]

\hrule
\begin{flalign}{\bf T1302912}: V/g^2 = -1.30291232,\;\mathfrak{so}(6)\to\emptyset&&\end{flalign}\\[-8.5ex]
\begin{flalign*}\begin{autobreak}m^2/m_0^2[\psi]: 
1.868_{\times 2},
2.939_{\times 2},
2.966_{\times 2},
3.715_{\times 2}\end{autobreak}&&\end{flalign*}\\[-8.5ex]
\begin{flalign*}\begin{autobreak}m^2/m_0^2[\phi]: 
-5.530,
-5.163,
-4.601,
-4.088,
-3.938,
-3.876,
-2.975,
-1.241,
-0.535,
0.000_{\times 18},
1.055,
10.323,
11.248,
11.304,
11.795,
12.471,
12.633,
15.614,
15.958,
16.392,
16.582,
22.633,
22.985,
23.633,
24.258\end{autobreak}&&\end{flalign*}\\[-8.5ex]
\begin{flalign*}\begin{autobreak}\kern0.0001em
\Lambda^1{}_1\approx-0.62113,
\Lambda^{2}{}_{2}\approx-0.58860,
\Lambda^{3}{}_{3}\approx0.24575,
\Lambda^{4}{}_{4}\approx0.29746,
\Lambda^{5}{}_{5}\approx0.32510,
\Lambda^{6}{}_{6}\approx0.34142\end{autobreak}&&\end{flalign*}\\[-8.5ex]
\begin{flalign*}\begin{autobreak}\kern0.0001em
\Sigma_{+123;2-456;1}\approx0.46285,
\Sigma_{+124;1-356;2}\approx0.38576,
\Sigma_{+126;2+345;1}\approx-0.06801,
\Sigma_{+135;1-246;2}\approx-0.73256,
\Sigma_{+145;2-236;1}\approx-0.31760,
\Sigma_{+156;1+234;2}\approx-0.44500\end{autobreak}&&\end{flalign*}\\[-8.5ex]
\\[3ex]

\hrule
\begin{flalign}{\bf T1304606}: V/g^2 = -1.30460644,\;\mathfrak{so}(6)\to\emptyset&&\end{flalign}\\[-8.5ex]
\begin{flalign*}\begin{autobreak}m^2/m_0^2[\psi]: 
1.863_{\times 2},
3.045_{\times 4},
3.623_{\times 2}\end{autobreak}&&\end{flalign*}\\[-8.5ex]
\begin{flalign*}\begin{autobreak}m^2/m_0^2[\phi]: 
-5.709,
-5.072,
-4.820,
-4.308,
-4.051,
-3.851,
-2.894,
-0.409,
0.000_{\times 18},
0.705_{\times 2},
10.707,
11.268,
12.495,
12.520_{\times 2},
13.495,
14.491,
15.601_{\times 2},
16.679,
22.729,
23.752_{\times 2},
24.989\end{autobreak}&&\end{flalign*}\\[-8.5ex]
\begin{flalign*}\begin{autobreak}\kern0.0001em
\Lambda^1{}_1=\Lambda^{2}{}_{2}\approx-0.60876,
\Lambda^{3}{}_{3}=\Lambda^{4}{}_{4}\approx0.28534,
\Lambda^{5}{}_{5}=\Lambda^{6}{}_{6}\approx0.32343\end{autobreak}&&\end{flalign*}\\[-8.5ex]
\begin{flalign*}\begin{autobreak}\kern0.0001em
\Sigma_{+123;2-456;1}=\Sigma_{+124;1-356;2}\approx-0.41059,
\Sigma_{+135;1-246;2}\approx0.32857,
\Sigma_{+145;2-236;1}\approx0.87485\end{autobreak}&&\end{flalign*}\\[-8.5ex]
\\[3ex]

\hrule
\begin{flalign}{\bf T1319179}: V/g^2 = -1.31917968,\;\mathfrak{so}(6)\to\emptyset&&\end{flalign}\\[-8.5ex]
\begin{flalign*}\begin{autobreak}m^2/m_0^2[\psi]: 
2.007_{\times 2},
2.314_{\times 2},
3.390_{\times 2},
3.479_{\times 2}\end{autobreak}&&\end{flalign*}\\[-8.5ex]
\begin{flalign*}\begin{autobreak}m^2/m_0^2[\phi]: 
-6.083_{\times 2},
-4.130,
-3.789_{\times 2},
-3.689,
-3.654_{\times 2},
-0.971,
0.000_{\times 17},
0.946_{\times 2},
8.035,
9.713,
10.759_{\times 2},
11.497,
12.191,
12.714,
17.658_{\times 2},
17.814,
21.132_{\times 2},
23.045,
23.374\end{autobreak}&&\end{flalign*}\\[-8.5ex]
\begin{flalign*}\begin{autobreak}\kern0.0001em
\Lambda^1{}_1=\Lambda^{2}{}_{2}\approx-0.60188,
\Lambda^{3}{}_{3}=\Lambda^{4}{}_{4}\approx0.22066,
\Lambda^{5}{}_{5}=\Lambda^{6}{}_{6}\approx0.38122\end{autobreak}&&\end{flalign*}\\[-8.5ex]
\begin{flalign*}\begin{autobreak}\kern0.0001em
\Sigma_{+123;2-456;1}=\Sigma_{+124;1-356;2}\approx0.59938,
\Sigma_{+135;1-246;2}\approx0.31540,
\Sigma_{+136;2-245;1}\approx-0.00424,
\Sigma_{+145;2-236;1}\approx0.24695,
\Sigma_{+146;1-235;2}\approx0.55810\end{autobreak}&&\end{flalign*}\\[-8.5ex]
\\[3ex]

\hrule
\begin{flalign}{\bf T1382251}: V/g^2 = -1.38225189,\;\mathfrak{so}(6)\to\emptyset&&\end{flalign}\\[-8.5ex]
\begin{flalign*}\begin{autobreak}m^2/m_0^2[\psi]: 
2.157_{\times 2},
2.626_{\times 2},
3.824_{\times 2},
4.425_{\times 2}\end{autobreak}&&\end{flalign*}\\[-8.5ex]
\begin{flalign*}\begin{autobreak}m^2/m_0^2[\phi]: 
-6.955,
-5.615,
-4.845,
-3.442,
-3.429,
-3.048,
-1.888,
-0.873,
0.000_{\times 18},
2.191,
4.190,
9.654,
9.868,
10.839,
11.089,
12.148,
14.269,
17.139,
17.449,
20.030,
20.351,
28.208,
28.346,
45.348,
45.357\end{autobreak}&&\end{flalign*}\\[-8.5ex]
\begin{flalign*}\begin{autobreak}\kern0.0001em
\Lambda^1{}_1=\Lambda^{2}{}_{2}\approx-0.61986,
\Lambda^{3}{}_{3}=\Lambda^{4}{}_{4}\approx0.12499,
\Lambda^{5}{}_{5}=\Lambda^{6}{}_{6}\approx0.49487\end{autobreak}&&\end{flalign*}\\[-8.5ex]
\begin{flalign*}\begin{autobreak}\kern0.0001em
\Sigma_{+123;1+456;2}=\Sigma_{+124;2+356;1}\approx-0.57846,
\Sigma_{+125;2-346;1}=\Sigma_{+126;1-345;2}\approx0.13899,
\Sigma_{+134;1+256;2}=\Sigma_{+156;1+234;2}\approx-0.44722,
\Sigma_{+135;1-246;2}\approx-0.37639,
\Sigma_{+136;2-245;1}\approx0.00529,
\Sigma_{+145;2-236;1}\approx0.14221,
\Sigma_{+146;1-235;2}\approx-0.12738\end{autobreak}&&\end{flalign*}\\[-8.5ex]
\\[3ex]

\hrule
\begin{flalign}{\bf T1391035}: V/g^2 = -1.39103566,\;\mathfrak{so}(6)\to\emptyset&&\end{flalign}\\[-8.5ex]
\begin{flalign*}\begin{autobreak}m^2/m_0^2[\psi]: 
2.576_{\times 4},
3.726_{\times 4}\end{autobreak}&&\end{flalign*}\\[-8.5ex]
\begin{flalign*}\begin{autobreak}m^2/m_0^2[\phi]: 
-7.422,
-5.873,
-5.133,
-3.439,
-3.258,
-3.039,
-1.551,
0.000_{\times 18},
0.990,
3.614,
5.111,
8.912,
9.520,
9.929,
11.031,
12.107,
15.840,
16.038,
16.254,
19.051,
20.981,
22.866,
23.627,
40.460,
40.498\end{autobreak}&&\end{flalign*}\\[-8.5ex]
\begin{flalign*}\begin{autobreak}\kern0.0001em
\Lambda^1{}_1\approx-0.65089,
\Lambda^{2}{}_{2}\approx-0.60760,
\Lambda^{3}{}_{3}\approx0.08094,
\Lambda^{4}{}_{4}\approx0.20776,
\Lambda^{5}{}_{5}\approx0.48464,
\Lambda^{6}{}_{6}\approx0.48515\end{autobreak}&&\end{flalign*}\\[-8.5ex]
\begin{flalign*}\begin{autobreak}\kern0.0001em
\Sigma_{+123;1+456;2}\approx0.61508,
\Sigma_{+124;2+356;1}\approx-0.57701,
\Sigma_{+126;1-345;2}\approx-0.20028,
\Sigma_{+134;2-256;1}\approx-0.56777,
\Sigma_{+136;1+245;2}\approx0.17976,
\Sigma_{+146;2+235;1}\approx0.41246\end{autobreak}&&\end{flalign*}\\[-8.5ex]
\\[3ex]

\hrule
\begin{flalign}{\bf T1416746}: V/g^2 = -1.41674647,\;\mathfrak{so}(6)\to\emptyset&&\end{flalign}\\[-8.5ex]
\begin{flalign*}\begin{autobreak}m^2/m_0^2[\psi]: 
2.541_{\times 4},
4.682_{\times 4}\end{autobreak}&&\end{flalign*}\\[-8.5ex]
\begin{flalign*}\begin{autobreak}m^2/m_0^2[\phi]: 
-7.588,
-6.331,
-4.664,
-3.637,
-2.868,
-2.381,
-0.500,
0.000_{\times 18},
2.018,
3.674,
5.078,
10.577,
11.395,
14.830,
15.208,
15.285,
17.454,
19.273,
19.969,
27.930,
27.963,
28.305,
28.585,
55.234,
55.252\end{autobreak}&&\end{flalign*}\\[-8.5ex]
\begin{flalign*}\begin{autobreak}\kern0.0001em
\Lambda^1{}_1\approx-0.73037,
\Lambda^{2}{}_{2}\approx-0.46700,
\Lambda^{3}{}_{3}\approx-0.03604,
\Lambda^{4}{}_{4}\approx0.22540,
\Lambda^{5}{}_{5}\approx0.50269,
\Lambda^{6}{}_{6}\approx0.50532\end{autobreak}&&\end{flalign*}\\[-8.5ex]
\begin{flalign*}\begin{autobreak}\kern0.0001em
\Sigma_{+123;1+456;2}\approx-0.63728,
\Sigma_{+126;1-345;2}\approx-0.22177,
\Sigma_{+134;2-256;1}\approx0.88862,
\Sigma_{+135;2+246;1}\approx-0.10297,
\Sigma_{+146;2+235;1}\approx0.35261,
\Sigma_{+156;2-234;1}\approx-0.20665\end{autobreak}&&\end{flalign*}\\[-8.5ex]
\\[3ex]

\hrule
\begin{flalign}{\bf T1417411}: V/g^2 = -1.41741118,\;\mathfrak{so}(6)\to\emptyset&&\end{flalign}\\[-8.5ex]
\begin{flalign*}\begin{autobreak}m^2/m_0^2[\psi]: 
2.667_{\times 4},
4.444_{\times 4}\end{autobreak}&&\end{flalign*}\\[-8.5ex]
\begin{flalign*}\begin{autobreak}m^2/m_0^2[\phi]: 
-7.719,
-6.560,
-4.692,
-3.535,
-2.851,
-2.492,
0.000_{\times 18},
0.276,
1.895,
4.301,
5.567,
10.235,
11.353,
14.431,
15.458,
15.502,
17.694,
20.814,
22.838,
24.048,
25.549,
25.724,
25.766,
53.186,
53.213\end{autobreak}&&\end{flalign*}\\[-8.5ex]
\begin{flalign*}\begin{autobreak}\kern0.0001em
\Lambda^1{}_1\approx-0.73052,
\Lambda^{2}{}_{2}\approx-0.47160,
\Lambda^{3}{}_{3}\approx-0.04027,
\Lambda^{4}{}_{4}\approx0.24055,
\Lambda^{5}{}_{5}\approx0.49948,
\Lambda^{6}{}_{6}\approx0.50237\end{autobreak}&&\end{flalign*}\\[-8.5ex]
\begin{flalign*}\begin{autobreak}\kern0.0001em
\Sigma_{+123;2-456;1}\approx-0.64026,
\Sigma_{+126;2+345;1}\approx0.23639,
\Sigma_{+134;1+256;2}\approx-0.90336,
\Sigma_{+146;1-235;2}\approx0.36681\end{autobreak}&&\end{flalign*}\\[-8.5ex]
\\[3ex]

\hrule
\begin{flalign}{\bf T1460654}: V/g^2 = -1.46065435,\;\mathfrak{so}(6)\to\emptyset&&\end{flalign}\\[-8.5ex]
\begin{flalign*}\begin{autobreak}m^2/m_0^2[\psi]: 
2.549_{\times 2},
2.872_{\times 2},
3.548_{\times 2},
4.727_{\times 2}\end{autobreak}&&\end{flalign*}\\[-8.5ex]
\begin{flalign*}\begin{autobreak}m^2/m_0^2[\phi]: 
-7.782,
-7.781,
-5.074,
-3.077_{\times 2},
-2.429,
-0.156,
0.000_{\times 18},
5.812_{\times 2},
8.206,
8.928,
9.049,
11.748,
15.607_{\times 2},
18.569,
18.574,
18.983,
23.132,
23.187,
26.681,
26.684,
44.562_{\times 2}\end{autobreak}&&\end{flalign*}\\[-8.5ex]
\begin{flalign*}\begin{autobreak}\kern0.0001em
\Lambda^1{}_1\approx-0.80626,
\Lambda^{2}{}_{2}\approx-0.17618,
\Lambda^{3}{}_{3}\approx0.04003,
\Lambda^{4}{}_{4}\approx0.04284,
\Lambda^{5}{}_{5}\approx0.44978,
\Lambda^{6}{}_{6}\approx0.44979\end{autobreak}&&\end{flalign*}\\[-8.5ex]
\begin{flalign*}\begin{autobreak}\kern0.0001em
\Sigma_{+123;2-456;1}\approx-0.83799,
\Sigma_{+124;1-356;2}\approx-0.83732,
\Sigma_{+125;2-346;1}\approx-0.03678,
\Sigma_{+126;1-345;2}\approx0.03663,
\Sigma_{+134;2-256;1}\approx-0.00494,
\Sigma_{+135;1-246;2}\approx-0.22033,
\Sigma_{+136;2-245;1}\approx-0.22065,
\Sigma_{+145;2-236;1}\approx-0.21718,
\Sigma_{+146;1-235;2}\approx0.21663,
\Sigma_{+156;2-234;1}\approx-0.13058\end{autobreak}&&\end{flalign*}\\[-8.5ex]
\\[3ex]

\hrule
\begin{flalign}{\bf T1460729}: V/g^2 = -1.46072960,\;\mathfrak{so}(6)\to\emptyset&&\end{flalign}\\[-8.5ex]
\begin{flalign*}\begin{autobreak}m^2/m_0^2[\psi]: 
2.578_{\times 2},
2.953_{\times 2},
3.433_{\times 2},
4.666_{\times 2}\end{autobreak}&&\end{flalign*}\\[-8.5ex]
\begin{flalign*}\begin{autobreak}m^2/m_0^2[\phi]: 
-7.837_{\times 2},
-5.119,
-3.105_{\times 2},
-2.400,
0.000_{\times 17},
0.081_{\times 2},
5.873_{\times 2},
8.042,
8.950_{\times 2},
11.735,
16.355_{\times 2},
17.388_{\times 2},
19.322,
23.288_{\times 2},
26.017_{\times 2},
43.861_{\times 2}\end{autobreak}&&\end{flalign*}\\[-8.5ex]
\begin{flalign*}\begin{autobreak}\kern0.0001em
\Lambda^1{}_1\approx-0.80503,
\Lambda^{2}{}_{2}\approx-0.18004,
\Lambda^{3}{}_{3}=\Lambda^{4}{}_{4}\approx0.04349,
\Lambda^{5}{}_{5}=\Lambda^{6}{}_{6}\approx0.44904\end{autobreak}&&\end{flalign*}\\[-8.5ex]
\begin{flalign*}\begin{autobreak}\kern0.0001em
\Sigma_{+123;1+456;2}\approx-0.81796,
\Sigma_{+123;2-456;1}=\Sigma_{+124;1-356;2}\approx-0.19809,
\Sigma_{+124;2+356;1}\approx0.81796,
\Sigma_{+135;1-246;2}=\Sigma_{+136;2-245;1}=\Sigma_{+145;2-236;1}\approx-0.15970,
\Sigma_{+135;2+246;1}\approx0.15381,
\Sigma_{+136;1+245;2}=\Sigma_{+145;1+236;2}=\Sigma_{+146;2+235;1}\approx-0.15381,
\Sigma_{+146;1-235;2}\approx0.15970\end{autobreak}&&\end{flalign*}\\[-8.5ex]
\\[3ex]

\hrule
\begin{flalign}{\bf T1497042}: V/g^2 = -1.49704248,\;\mathfrak{so}(6)\to\emptyset&&\end{flalign}\\[-8.5ex]
\begin{flalign*}\begin{autobreak}m^2/m_0^2[\psi]: 
2.564_{\times 2},
2.744_{\times 2},
4.924_{\times 2},
4.987_{\times 2}\end{autobreak}&&\end{flalign*}\\[-8.5ex]
\begin{flalign*}\begin{autobreak}m^2/m_0^2[\phi]: 
-7.876,
-7.552,
-6.086,
-2.529,
-1.033,
-0.827,
0.000_{\times 18},
3.290,
4.630,
5.745,
7.840,
8.291,
9.757,
11.599,
15.632,
18.415,
18.640,
21.703,
21.866,
28.340,
28.422,
33.168,
33.169,
60.129,
60.135\end{autobreak}&&\end{flalign*}\\[-8.5ex]
\begin{flalign*}\begin{autobreak}\kern0.0001em
\Lambda^1{}_1\approx-0.82463,
\Lambda^{2}{}_{2}\approx-0.19891,
\Lambda^{3}{}_{3}\approx0.00718,
\Lambda^{4}{}_{4}\approx0.00932,
\Lambda^{5}{}_{5}\approx0.50354,
\Lambda^{6}{}_{6}\approx0.50351\end{autobreak}&&\end{flalign*}\\[-8.5ex]
\begin{flalign*}\begin{autobreak}\kern0.0001em
\Sigma_{+123;2-456;1}\approx0.82383,
\Sigma_{+124;1-356;2}\approx0.81305,
\Sigma_{+125;1+346;2}\approx-0.13392,
\Sigma_{+136;1+245;2}\approx-0.28798,
\Sigma_{+146;2+235;1}\approx-0.25309,
\Sigma_{+156;2-234;1}\approx0.38869\end{autobreak}&&\end{flalign*}\\[-8.5ex]
\\[3ex]

\hrule
\begin{flalign}{\bf T1499666}: V/g^2 = -1.49966681,\;\mathfrak{so}(6)\to\emptyset&&\end{flalign}\\[-8.5ex]
\begin{flalign*}\begin{autobreak}m^2/m_0^2[\psi]: 
2.796_{\times 4},
4.671_{\times 4}\end{autobreak}&&\end{flalign*}\\[-8.5ex]
\begin{flalign*}\begin{autobreak}m^2/m_0^2[\phi]: 
-8.146,
-7.859,
-6.341,
-2.381,
-0.858,
0.000_{\times 18},
0.587,
3.588,
4.575,
6.215,
7.213,
8.309,
8.389,
11.542,
16.667,
18.400,
18.880,
22.379,
22.570,
25.467,
25.509,
31.179,
31.185,
57.469,
57.478\end{autobreak}&&\end{flalign*}\\[-8.5ex]
\begin{flalign*}\begin{autobreak}\kern0.0001em
\Lambda^1{}_1\approx-0.81651,
\Lambda^{2}{}_{2}\approx-0.22484,
\Lambda^{3}{}_{3}\approx-0.00212,
\Lambda^{4}{}_{4}\approx0.03391,
\Lambda^{5}{}_{5}\approx0.50465,
\Lambda^{6}{}_{6}\approx0.50490\end{autobreak}&&\end{flalign*}\\[-8.5ex]
\begin{flalign*}\begin{autobreak}\kern0.0001em
\Sigma_{+123;2-456;1}\approx0.83606,
\Sigma_{+124;1-356;2}\approx0.83978,
\Sigma_{+126;2+345;1}\approx-0.09844,
\Sigma_{+135;2+246;1}\approx-0.29598,
\Sigma_{+145;1+236;2}\approx0.27814,
\Sigma_{+156;2-234;1}\approx-0.26863\end{autobreak}&&\end{flalign*}\\[-8.5ex]
\\[3ex]

\hrule
\begin{flalign}{\bf T1501862}: V/g^2 = -1.50186250,\;\mathfrak{so}(6)\to\emptyset&&\end{flalign}\\[-8.5ex]
\begin{flalign*}\begin{autobreak}m^2/m_0^2[\psi]: 
2.958_{\times 4},
4.361_{\times 4}\end{autobreak}&&\end{flalign*}\\[-8.5ex]
\begin{flalign*}\begin{autobreak}m^2/m_0^2[\phi]: 
-8.390,
-8.149,
-6.547,
-2.243,
0.000_{\times 18},
0.534_{\times 2},
3.715,
4.659,
6.331,
6.591,
8.036_{\times 2},
11.530,
17.908,
19.043,
19.546,
22.488_{\times 2},
22.827_{\times 2},
28.910,
28.950,
54.551,
54.564\end{autobreak}&&\end{flalign*}\\[-8.5ex]
\begin{flalign*}\begin{autobreak}\kern0.0001em
\Lambda^1{}_1\approx-0.81048,
\Lambda^{2}{}_{2}\approx-0.24631,
\Lambda^{3}{}_{3}=\Lambda^{4}{}_{4}\approx0.02461,
\Lambda^{5}{}_{5}\approx0.50349,
\Lambda^{6}{}_{6}\approx0.50407\end{autobreak}&&\end{flalign*}\\[-8.5ex]
\begin{flalign*}\begin{autobreak}\kern0.0001em
\Sigma_{+123;1+456;2}\approx0.85496,
\Sigma_{+124;2+356;1}\approx-0.85496,
\Sigma_{+135;1-246;2}=\Sigma_{+145;2-236;1}\approx0.30382\end{autobreak}&&\end{flalign*}\\[-8.5ex]
\\[3ex]

\hrule
\begin{flalign}{\bf T1510900}: V/g^2 = -1.51090053,\;\mathfrak{so}(6)\to\emptyset&&\end{flalign}\\[-8.5ex]
\begin{flalign*}\begin{autobreak}m^2/m_0^2[\psi]: 
2.341_{\times 2},
2.515_{\times 2},
5.252_{\times 2},
5.535_{\times 2}\end{autobreak}&&\end{flalign*}\\[-8.5ex]
\begin{flalign*}\begin{autobreak}m^2/m_0^2[\phi]: 
-7.096,
-6.092,
-5.664,
-3.912,
-3.831,
-1.977,
0.000_{\times 18},
4.992,
5.327,
5.471,
8.192,
10.978,
11.323,
12.154,
15.682,
18.760,
19.789,
23.119,
24.250,
32.619,
32.622,
38.729,
38.866,
54.809,
54.833\end{autobreak}&&\end{flalign*}\\[-8.5ex]
\begin{flalign*}\begin{autobreak}\kern0.0001em
\Lambda^1{}_1\approx-0.80362,
\Lambda^{2}{}_{2}\approx-0.54087,
\Lambda^{3}{}_{3}\approx0.11891,
\Lambda^{4}{}_{4}\approx0.28645,
\Lambda^{5}{}_{5}\approx0.46652,
\Lambda^{6}{}_{6}\approx0.47261\end{autobreak}&&\end{flalign*}\\[-8.5ex]
\begin{flalign*}\begin{autobreak}\kern0.0001em
\Sigma_{+123;2-456;1}\approx-0.65751,
\Sigma_{+126;1-345;2}\approx-0.15110,
\Sigma_{+134;1+256;2}\approx-0.55854,
\Sigma_{+135;2+246;1}\approx0.32344,
\Sigma_{+146;2+235;1}\approx0.35501,
\Sigma_{+156;1+234;2}\approx-0.74695\end{autobreak}&&\end{flalign*}\\[-8.5ex]
\\[3ex]

\hrule
\begin{flalign}{\bf T1547778}: V/g^2 = -1.54777840,\;\mathfrak{so}(6)\to\emptyset&&\end{flalign}\\[-8.5ex]
\begin{flalign*}\begin{autobreak}m^2/m_0^2[\psi]: 
2.919_{\times 2},
3.511_{\times 4},
4.696_{\times 2}\end{autobreak}&&\end{flalign*}\\[-8.5ex]
\begin{flalign*}\begin{autobreak}m^2/m_0^2[\phi]: 
-7.690,
-7.121,
-5.196,
-3.768,
-3.229,
0.000_{\times 18},
2.491,
5.758,
5.882,
7.906_{\times 2},
11.876,
15.646,
16.511,
16.875_{\times 2},
18.734,
19.897,
21.602_{\times 2},
25.828,
34.397,
37.884_{\times 2},
40.118\end{autobreak}&&\end{flalign*}\\[-8.5ex]
\begin{flalign*}\begin{autobreak}\kern0.0001em
\Lambda^1{}_1=\Lambda^{2}{}_{2}\approx-0.71461,
\Lambda^{3}{}_{3}\approx0.13652,
\Lambda^{4}{}_{4}=\Lambda^{5}{}_{5}\approx0.41599,
\Lambda^{6}{}_{6}\approx0.46074\end{autobreak}&&\end{flalign*}\\[-8.5ex]
\begin{flalign*}\begin{autobreak}\kern0.0001em
\Sigma_{+123;1+456;2}\approx0.70763,
\Sigma_{+126;2+345;1}\approx-0.25491,
\Sigma_{+134;1+256;2}=\Sigma_{+146;2+235;1}\approx0.45958,
\Sigma_{+135;2+246;1}=\Sigma_{+156;1+234;2}\approx-0.51427\end{autobreak}&&\end{flalign*}\\[-8.5ex]
\\[3ex]

\hrule
\begin{flalign}{\bf T1738407}: V/g^2 = -1.73840792,\;\mathfrak{so}(6)\to\emptyset&&\end{flalign}\\[-8.5ex]
\begin{flalign*}\begin{autobreak}m^2/m_0^2[\psi]: 
2.380_{\times 2},
3.106_{\times 2},
5.205_{\times 2},
5.322_{\times 2}\end{autobreak}&&\end{flalign*}\\[-8.5ex]
\begin{flalign*}\begin{autobreak}m^2/m_0^2[\phi]: 
-6.504_{\times 2},
-6.250_{\times 2},
-3.019,
0.000_{\times 17},
3.819_{\times 2},
11.175,
11.214_{\times 2},
13.297,
17.275_{\times 2},
21.000_{\times 2},
25.218_{\times 2},
26.699_{\times 2},
28.635,
30.545_{\times 2},
34.291_{\times 2},
35.846\end{autobreak}&&\end{flalign*}\\[-8.5ex]
\begin{flalign*}\begin{autobreak}\kern0.0001em
\Lambda^1{}_1\approx-1.11580,
\Lambda^{2}{}_{2}\approx0.20510,
\Lambda^{3}{}_{3}=\Lambda^{4}{}_{4}\approx0.22634,
\Lambda^{5}{}_{5}=\Lambda^{6}{}_{6}\approx0.22901\end{autobreak}&&\end{flalign*}\\[-8.5ex]
\begin{flalign*}\begin{autobreak}\kern0.0001em
\Sigma_{+125;1+346;2}=\Sigma_{+126;2+345;1}\approx-0.80729,
\Sigma_{+135;1-246;2}=\Sigma_{+145;2-236;1}=\Sigma_{+146;1-235;2}\approx0.42803,
\Sigma_{+136;2-245;1}\approx-0.42803\end{autobreak}&&\end{flalign*}\\[-8.5ex]
\\[3ex]

\hrule\smallskip\hrule

\section{Scalar mass spectra for the classic vacua}
\label{appendixC}

\begin{table}[t]
\renewcommand{\arraystretch}{0.9}
\begin{minipage}{0.48\linewidth}
\begin{minipage}{1\linewidth}
\centering
\scalebox{0.9}{
\begin{tabular}{@{\extracolsep{10 pt}} c c  c }
\toprule
\noalign{\smallskip}
Degeneracy &  $m^2L^2$    & BF Stability  \\
\noalign{\smallskip}
\midrule
\noalign{\medskip}
2& 0 &    S   \\ 
20& $-3$ &    S   \\ 
20& $-4$ &    S   \\  
\noalign{\smallskip}
\bottomrule
\end{tabular}
}
\caption{\label{tbl:SO6spec} The $\SO(6)$ point.}
\end{minipage}
\bigskip 

\begin{minipage}{1\linewidth}\centering
\scalebox{0.9}{
\begin{tabular}{@{\extracolsep{10 pt}} c c  c }
\toprule
\noalign{\smallskip}
Degeneracy &  $m^2L^2$    & Stability  \\
\noalign{\smallskip}
\midrule
\noalign{\medskip}
1& 8 &    S   \\ 
7& $0$ &    S   \\ 
20& $-2$ &    S   \\ 
14& $-16/3$ &    U   \\   
\noalign{\smallskip}
\bottomrule
\end{tabular}
}
\caption{\label{tbl:SO5spec} The  $\SO(5)$ point.}
\end{minipage}
\bigskip

\begin{minipage}{1\linewidth}\centering
\scalebox{0.9}{
\begin{tabular}{@{\extracolsep{10 pt}} c c  c }
\toprule
\noalign{\smallskip}
Degeneracy &  $m^2L^2$    & Stability  \\
\noalign{\smallskip}
\midrule
\noalign{\medskip}
1& 8 &    S   \\ 
17& $0$ &    S   \\ 
12& $-16/9$ &    S   \\ 
12& $-40/9$ &    U   \\  
\noalign{\smallskip}
\bottomrule
\end{tabular}
}
\caption{\label{tbl:SU3spec} The $\SU(3)$ point.}
\end{minipage}
\end{minipage}
\begin{minipage}{0.48\linewidth}
\begin{minipage}{1\linewidth}\centering
\scalebox{0.9}{
\begin{tabular}{@{\extracolsep{10 pt}} c c  c }
\toprule
\noalign{\smallskip}
Degeneracy &  $m^2L^2$    & BF stability  \\
\noalign{\smallskip}
\midrule
\noalign{\medskip}
1& $4+2\sqrt{7}$ &    S   \\ 
2& $3$ &    S   \\ 
13& $0$ &    S   \\  
1& $4-2\sqrt{7}$ &    S   \\  
4& $-39/16$ &    S   \\ 
2& $-3$ &    S   \\ 
4& $-55/16$ &    S   \\ 
12& $-15/4$ &    S   \\ 
3& $-4$ &    S   \\ 
\noalign{\smallskip}
\bottomrule
\end{tabular}
}
\caption{\label{tbl:SU2U1spec} The $\SU(2)\times \U(1)$ point.}
\end{minipage}
\begin{minipage}{1\linewidth}\centering
\bigskip
\scalebox{0.9}{
\begin{tabular}{@{\extracolsep{10 pt}} c c  c }
\toprule
\noalign{\smallskip}
Degeneracy &  $m^2L^2$    & Stability  \\
\noalign{\smallskip}
\midrule
\noalign{\medskip}
2& $\fracc{52}{5}$ &    S   \\ 
1& $\fracc{48}{5}$ &    S   \\ 
2& $\fracc{84}{25}$ &    S   \\ 
13& $0$ &    S   \\ 
6& $-\fracc{12}{5}$ &    S   \\ 
8& $-\fracc{64}{25}$ &    S   \\ 
4& $-4$ &    S   \\  
6& $-\fracc{136}{25}$ &    U   \\ 
\noalign{\smallskip}
\bottomrule
\end{tabular}
}
\caption{\label{tbl:SU2U1U1spec} The $\SU(2)\times \U(1)^2$ point.}
\end{minipage}
\end{minipage}
\end{table}

In this appendix we collect results for the spectrum of scalar fluctuations around each of the five classic critical points found in \cite{Khavaev:1998fb}. 

The spectrum of scalar masses at the $\SO(6)$ point, {\tt T075000}, in Table~\ref{tbl:SO6spec} follows from  $\mathcal{N}=8$ supersymmetry. At the $\SU(2)\times \U(1)$ point,  {\tt T0839947},  the full spectrum was computed and organized into multiplets of $\mathcal{N}=2$ supersymmetry  in \cite{Freedman:1999gp} with the scalar masses  given in Table~\ref{tbl:SU2U1spec}.
Both points are perturbatively stable. The BF instability of the  $\SU(3)$ point,  {\tt T0843750}, was established in \cite{Freedman:1999dz} and subsequently confirmed in \cite{Distler:1999tr,Girardello:1998pd}. The full scalar spectrum at this point, see also \cite{Distler:1998gb}, is given in Table~\ref{tbl:SU3spec}. Finally, the scalar spectra at the 
 the $\SO(5)$ point, \T0780031, and the $\SU(2)\times \U(1)^2$ point, \T0870298, were computed in  1999  \cite{Pilch:1999misc} and are given in Tables~\ref{tbl:SO5spec} and \ref{tbl:SU2U1U1spec}.


\bibliography{5dN8}

\providecommand{\href}[2]{#2}\begingroup\raggedright\begin{thebibliography}{10}

\bibitem{Gunaydin:1984qu}
M.~Gunaydin, L.~J. Romans, and N.~P. Warner, {\it {Gauged N=8 Supergravity in
  Five-Dimensions}},  {\em Phys. Lett.} {\bf 154B} (1985) 268--274.

\bibitem{Gunaydin:1985cu}
M.~Gunaydin, L.~J. Romans, and N.~P. Warner, {\it {Compact and Noncompact
  Gauged Supergravity Theories in Five-Dimensions}},  {\em Nucl. Phys.} {\bf
  B272} (1986) 598--646.

\bibitem{Pernici:1985ju}
M.~Pernici, K.~Pilch, and P.~van Nieuwenhuizen, {\it {Gauged N=8 D=5
  Supergravity}},  {\em Nucl. Phys.} {\bf B259} (1985) 460.

\bibitem{Khavaev:1998fb}
A.~Khavaev, K.~Pilch, and N.~P. Warner, {\it {New vacua of gauged N=8
  supergravity in five-dimensions}},  {\em Phys. Lett.} {\bf B487} (2000)
  14--21, [\href{http://arxiv.org/abs/hep-th/9812035}{{\tt hep-th/9812035}}].

\bibitem{Cvetic:2000nc}
M.~Cvetic, H.~Lu, C.~N. Pope, A.~Sadrzadeh, and T.~A. Tran, {\it {Consistent
  SO(6) reduction of type IIB supergravity on S**5}},  {\em Nucl. Phys.} {\bf
  B586} (2000) 275--286, [\href{http://arxiv.org/abs/hep-th/0003103}{{\tt
  hep-th/0003103}}].

\bibitem{Pilch:2000ue}
K.~Pilch and N.~P. Warner, {\it {N=2 supersymmetric RG flows and the IIB
  dilaton}},  {\em Nucl. Phys.} {\bf B594} (2001) 209--228,
  [\href{http://arxiv.org/abs/hep-th/0004063}{{\tt hep-th/0004063}}].

\bibitem{Lee:2014mla}
K.~Lee, C.~Strickland-Constable, and D.~Waldram, {\it {Spheres, generalised
  parallelisability and consistent truncations}},  {\em Fortsch. Phys.} {\bf
  65} (2017), no.~10-11 1700048, [\href{http://arxiv.org/abs/1401.3360}{{\tt
  arXiv:1401.3360}}].

\bibitem{Baguet:2015sma}
A.~Baguet, O.~Hohm, and H.~Samtleben, {\it {Consistent Type IIB Reductions to
  Maximal 5D Supergravity}},  {\em Phys. Rev.} {\bf D92} (2015), no.~6 065004,
  [\href{http://arxiv.org/abs/1506.01385}{{\tt arXiv:1506.01385}}].

\bibitem{Comsa:2019rcz}
I.~M. Comsa, M.~Firsching, and T.~Fischbacher, {\it {SO(8) Supergravity and the
  Magic of Machine Learning}},  \href{http://arxiv.org/abs/1906.00207}{{\tt
  arXiv:1906.00207}}.

\bibitem{deWit:1982bul}
B.~de~Wit and H.~Nicolai, {\it {N=8 Supergravity}},  {\em Nucl. Phys.} {\bf
  B208} (1982) 323.

\bibitem{Bobev:2019dik}
N.~Bobev, T.~Fischbacher, and K.~Pilch, {\it {Properties of the new $
  \mathcal{N} $ = 1 AdS$_{4}$ vacuum of maximal supergravity}},  {\em JHEP}
  {\bf 01} (2020) 099, [\href{http://arxiv.org/abs/1909.10969}{{\tt
  arXiv:1909.10969}}].

\bibitem{Girardello:1998pd}
L.~Girardello, M.~Petrini, M.~Porrati, and A.~Zaffaroni, {\it {Novel local CFT
  and exact results on perturbations of N=4 superYang Mills from AdS
  dynamics}},  {\em JHEP} {\bf 12} (1998) 022,
  [\href{http://arxiv.org/abs/hep-th/9810126}{{\tt hep-th/9810126}}].

\bibitem{Distler:1998gb}
J.~Distler and F.~Zamora, {\it {Nonsupersymmetric conformal field theories from
  stable anti-de Sitter spaces}},  {\em Adv. Theor. Math. Phys.} {\bf 2} (1999)
  1405--1439, [\href{http://arxiv.org/abs/hep-th/9810206}{{\tt
  hep-th/9810206}}].

\bibitem{Freedman:1999gp}
D.~Z. Freedman, S.~S. Gubser, K.~Pilch, and N.~P. Warner, {\it {Renormalization
  group flows from holography supersymmetry and a c theorem}},  {\em Adv.
  Theor. Math. Phys.} {\bf 3} (1999) 363--417,
  [\href{http://arxiv.org/abs/hep-th/9904017}{{\tt hep-th/9904017}}].

\bibitem{Gibbons:1983aq}
G.~W. Gibbons, C.~M. Hull, and N.~P. Warner, {\it {The Stability of Gauged
  Supergravity}},  {\em Nucl. Phys.} {\bf B218} (1983) 173.

\bibitem{Breitenlohner:1982jf}
P.~Breitenlohner and D.~Z. Freedman, {\it {Stability in Gauged Extended
  Supergravity}},  {\em Annals Phys.} {\bf 144} (1982) 249.

\bibitem{Ooguri:2016pdq}
H.~Ooguri and C.~Vafa, {\it {Non-supersymmetric AdS and the Swampland}},  {\em
  Adv. Theor. Math. Phys.} {\bf 21} (2017) 1787--1801,
  [\href{http://arxiv.org/abs/1610.01533}{{\tt arXiv:1610.01533}}].

\bibitem{ArkaniHamed:2006dz}
N.~Arkani-Hamed, L.~Motl, A.~Nicolis, and C.~Vafa, {\it {The String landscape,
  black holes and gravity as the weakest force}},  {\em JHEP} {\bf 06} (2007)
  060, [\href{http://arxiv.org/abs/hep-th/0601001}{{\tt hep-th/0601001}}].

\bibitem{Fischbacher:2011jx}
T.~Fischbacher, {\it {The Encyclopedic Reference of Critical Points for
  SO(8)-Gauged N=8 Supergravity. Part 1: Cosmological Constants in the Range
  -$\Lambda/g^2 \in [6:14.7)$}},  \href{http://arxiv.org/abs/1109.1424}{{\tt
  arXiv:1109.1424}}.

\bibitem{Leigh:1995ep}
R.~G. Leigh and M.~J. Strassler, {\it {Exactly marginal operators and duality
  in four-dimensional N=1 supersymmetric gauge theory}},  {\em Nucl. Phys.}
  {\bf B447} (1995) 95--136, [\href{http://arxiv.org/abs/hep-th/9503121}{{\tt
  hep-th/9503121}}].

\bibitem{Warner:1983du}
N.~P. Warner, {\it {Some Properties of the Scalar Potential in Gauged
  Supergravity Theories}},  {\em Nucl. Phys.} {\bf B231} (1984) 250--268.

\bibitem{Fischbacher:2010ec}
T.~Fischbacher, K.~Pilch, and N.~P. Warner, {\it {New Supersymmetric and
  Stable, Non-Supersymmetric Phases in Supergravity and Holographic Field
  Theory}},  \href{http://arxiv.org/abs/1010.4910}{{\tt arXiv:1010.4910}}.

\bibitem{Bena:2020misc}
I.~Bena, K.~Pilch, and N.~P. Warner, {\it {Brane-Jet Instabilities}},
  \href{http://arxiv.org/abs/2003.02851}{{\tt arXiv:2003.02851}}.

\bibitem{Malek:2020misc}
E.~Malek, H.~Nicolai, and H.~Samtleben. In preparation, 2020.

\bibitem{Fischbacher:2002fx}
T.~Fischbacher, H.~Nicolai, and H.~Samtleben, {\it {Vacua of maximal gauged D =
  3 supergravities}},  {\em Class. Quant. Grav.} {\bf 19} (2002) 5297--5334,
  [\href{http://arxiv.org/abs/hep-th/0207206}{{\tt hep-th/0207206}}].

\bibitem{Fischbacher:2008zu}
T.~Fischbacher, {\it {The Many vacua of gauged extended supergravities}},  {\em
  Gen. Rel. Grav.} {\bf 41} (2009) 315--411,
  [\href{http://arxiv.org/abs/0811.1915}{{\tt arXiv:0811.1915}}].

\bibitem{Warner:1983vz}
N.~P. Warner, {\it {Some New Extrema of the Scalar Potential of Gauged $N=8$
  Supergravity}},  {\em Phys. Lett.} {\bf 128B} (1983) 169--173.

\bibitem{Fischbacher:2009cj}
T.~Fischbacher, {\it {Fourteen new stationary points in the scalar potential of
  SO(8)-gauged N=8, D=4 supergravity}},  {\em JHEP} {\bf 09} (2010) 068,
  [\href{http://arxiv.org/abs/0912.1636}{{\tt arXiv:0912.1636}}].

\bibitem{Borghese:2013dja}
A.~Borghese, A.~Guarino, and D.~Roest, {\it {Triality, Periodicity and
  Stability of SO(8) Gauged Supergravity}},  {\em JHEP} {\bf 05} (2013) 107,
  [\href{http://arxiv.org/abs/1302.6057}{{\tt arXiv:1302.6057}}].

\bibitem{Pilch:2000fu}
K.~Pilch and N.~P. Warner, {\it {N=1 supersymmetric renormalization group flows
  from IIB supergravity}},  {\em Adv. Theor. Math. Phys.} {\bf 4} (2002)
  627--677, [\href{http://arxiv.org/abs/hep-th/0006066}{{\tt hep-th/0006066}}].

\bibitem{Bobev:2016nua}
N.~Bobev, H.~Elvang, U.~Kol, T.~Olson, and S.~S. Pufu, {\it {Holography for $
  \mathcal{N} $ = 1$^{?}$ on S$^{4}$}},  {\em JHEP} {\bf 10} (2016) 095,
  [\href{http://arxiv.org/abs/1605.00656}{{\tt arXiv:1605.00656}}].

\bibitem{Fischbacher:2010ki}
T.~Fischbacher, {\it {Numerical tools to validate stationary points of
  SO(8)-gauged N=8 D=4 supergravity}},  {\em Comput. Phys. Commun.} {\bf 183}
  (2012) 780--784, [\href{http://arxiv.org/abs/1007.0600}{{\tt
  arXiv:1007.0600}}].

\bibitem{Abadi2016}
M.~Abadi, P.~Barham, J.~Chen, Z.~Chen, A.~Davis, J.~Dean, M.~Devin,
  S.~Ghemawat, G.~Irving, M.~Isard, M.~Kudlur, J.~Levenberg, R.~Monga,
  S.~Moore, D.~G. Murray, B.~Steiner, P.~Tucker, V.~Vasudevan, P.~Warden,
  M.~Wicke, Y.~Yu, and X.~Zheng, {\it {TensorFlow: A system for large-scale
  machine learning}},  in {\em 12th USENIX Symposium on Operating Systems
  Design and Implementation (OSDI 16)}, pp.~265--283, 2016.

\bibitem{Khavaev:2000gb}
A.~Khavaev and N.~P. Warner, {\it {A Class of N=1 supersymmetric RG flows from
  five-dimensional N=8 supergravity}},  {\em Phys. Lett.} {\bf B495} (2000)
  215--222, [\href{http://arxiv.org/abs/hep-th/0009159}{{\tt hep-th/0009159}}].

\bibitem{Andrianopoli:1996bq}
L.~Andrianopoli, R.~D'Auria, S.~Ferrara, P.~Fre, and M.~Trigiante, {\it {RR
  scalars, U duality and solvable Lie algebras}},  {\em Nucl. Phys. B} {\bf
  496} (1997) 617--629, [\href{http://arxiv.org/abs/hep-th/9611014}{{\tt
  hep-th/9611014}}].

\bibitem{Andrianopoli:1996zg}
L.~Andrianopoli, R.~D'Auria, S.~Ferrara, P.~Fre, R.~Minasian, and M.~Trigiante,
  {\it {Solvable Lie algebras in type IIA, type IIB and M theories}},  {\em
  Nucl. Phys. B} {\bf 493} (1997) 249--280,
  [\href{http://arxiv.org/abs/hep-th/9612202}{{\tt hep-th/9612202}}].

\bibitem{FTtalks2020}
T.~Fischbacher, ``{\it Studying M-Theory Spontaneous Symmetry Breaking with
  Machine Learning Tools}.'' Seminar at
  \href{https://theorycolloquium.phys.ethz.ch/programme/previous/autumn19.html}{ETH,
  Zurich, November 4, 2019.} Seminar at
  \href{https://www.gravity.physik.fau.de/calendar/studying-m-theory-spontaneous-symmetry-breaking-with-machine-learning-tools/}{Erlangen
  University, November 27, 2019}.

\bibitem{KPtalks2020}
K.~Pilch, ``{\it AdS vacua of maximal supergravities}.'' Seminar at
  \href{http://semparis.lpthe.jussieu.fr/poster?searchpattern=13555&searchfield=Key&outputform=3}{IPhT
  Saclay, February 5, 2020.} Lecture at the
  \href{https://southweststringsme.wixsite.com/website/program}{Southwest
  Strings Meeting 2020, Utah State University, February 14-15, 2020}.

\bibitem{Krishnan:2020sfg}
C.~Krishnan, V.~Mohan, and S.~Ray, {\it {Machine Learning ${\cal N}=8, D=5$
  Gauged Supergravity}},  \href{http://arxiv.org/abs/2002.12927}{{\tt
  arXiv:2002.12927}}.

\bibitem{Girardello:1999bd}
L.~Girardello, M.~Petrini, M.~Porrati, and A.~Zaffaroni, {\it {The Supergravity
  dual of N=1 superYang-Mills theory}},  {\em Nucl. Phys.} {\bf B569} (2000)
  451--469, [\href{http://arxiv.org/abs/hep-th/9909047}{{\tt hep-th/9909047}}].

\bibitem{Bobev:2010de}
N.~Bobev, A.~Kundu, K.~Pilch, and N.~P. Warner, {\it {Supersymmetric Charged
  Clouds in AdS$_5$}},  {\em JHEP} {\bf 03} (2011) 070,
  [\href{http://arxiv.org/abs/1005.3552}{{\tt arXiv:1005.3552}}].

\bibitem{Aprile:2011uq}
F.~Aprile, D.~Roest, and J.~G. Russo, {\it {Holographic Superconductors from
  Gauged Supergravity}},  {\em JHEP} {\bf 06} (2011) 040,
  [\href{http://arxiv.org/abs/1104.4473}{{\tt arXiv:1104.4473}}].

\bibitem{Bobev:2013cja}
N.~Bobev, H.~Elvang, D.~Z. Freedman, and S.~S. Pufu, {\it {Holography for $N =
  2^*$ on $S^4$}},  {\em JHEP} {\bf 07} (2014) 001,
  [\href{http://arxiv.org/abs/1311.1508}{{\tt arXiv:1311.1508}}].

\bibitem{Bobev:2014jva}
N.~Bobev, K.~Pilch, and O.~Vasilakis, {\it {(0, 2) SCFTs from the
  Leigh-Strassler fixed point}},  {\em JHEP} {\bf 06} (2014) 094,
  [\href{http://arxiv.org/abs/1403.7131}{{\tt arXiv:1403.7131}}].

\bibitem{DallAgata:2011aa}
G.~Dall'Agata and G.~Inverso, {\it {On the Vacua of N = 8 Gauged Supergravity
  in 4 Dimensions}},  {\em Nucl. Phys.} {\bf B859} (2012) 70--95,
  [\href{http://arxiv.org/abs/1112.3345}{{\tt arXiv:1112.3345}}].

\bibitem{jax2018github}
J.~Bradbury, R.~Frostig, P.~Hawkins, M.~J. Johnson, C.~Leary, D.~Maclaurin, and
  S.~Wanderman-Milne, {\it {JAX}: composable transformations of
  {P}ython+{N}um{P}y programs, v 0.1.55},  2018.
\newblock \href{http://github.com/google/jax}{http://github.com/google/jax}.

\bibitem{helgason1979differential}
S.~Helgason, {\em Differential Geometry, Lie Groups, and Symmetric Spaces}.
\newblock ISSN. Elsevier Science, 1979.

\bibitem{Bianchi:2000sm}
M.~Bianchi, O.~DeWolfe, D.~Z. Freedman, and K.~Pilch, {\it {Anatomy of two
  holographic renormalization group flows}},  {\em JHEP} {\bf 01} (2001) 021,
  [\href{http://arxiv.org/abs/hep-th/0009156}{{\tt hep-th/0009156}}].

\bibitem{Gorbenko:2018ncu}
V.~Gorbenko, S.~Rychkov, and B.~Zan, {\it {Walking, Weak first-order
  transitions, and Complex CFTs}},  {\em JHEP} {\bf 10} (2018) 108,
  [\href{http://arxiv.org/abs/1807.11512}{{\tt arXiv:1807.11512}}].

\bibitem{Faedo:2019nxw}
A.~F. Faedo, C.~Hoyos, D.~Mateos, and J.~G. Subils, {\it {Holographic Complex
  CFTs}},  \href{http://arxiv.org/abs/1909.04008}{{\tt arXiv:1909.04008}}.

\bibitem{Donos:2017sba}
A.~Donos, J.~P. Gauntlett, C.~Rosen, and O.~Sosa-Rodriguez, {\it {Boomerang RG
  flows with intermediate conformal invariance}},  {\em JHEP} {\bf 04} (2018)
  017, [\href{http://arxiv.org/abs/1712.08017}{{\tt arXiv:1712.08017}}].

\bibitem{Bobev:2019wnf}
N.~Bobev, F.~F. Gautason, B.~E. Niehoff, and J.~van Muiden, {\it {A holographic
  kaleidoscope for $ \mathcal{N} $ = 1*}},  {\em JHEP} {\bf 10} (2019) 185,
  [\href{http://arxiv.org/abs/1906.09270}{{\tt arXiv:1906.09270}}].

\bibitem{Bobev:2010ib}
N.~Bobev, N.~Halmagyi, K.~Pilch, and N.~P. Warner, {\it {Supergravity
  Instabilities of Non-Supersymmetric Quantum Critical Points}},  {\em Class.
  Quant. Grav.} {\bf 27} (2010) 235013,
  [\href{http://arxiv.org/abs/1006.2546}{{\tt arXiv:1006.2546}}].

\bibitem{deWit:2002vt}
B.~de~Wit, H.~Samtleben, and M.~Trigiante, {\it {On Lagrangians and gaugings of
  maximal supergravities}},  {\em Nucl. Phys.} {\bf B655} (2003) 93--126,
  [\href{http://arxiv.org/abs/hep-th/0212239}{{\tt hep-th/0212239}}].

\bibitem{Freedman:1999dz}
D.~Z. Freedman, S.~S. Gubser, K.~Pilch, and N.~P. Warner, ``{Private
  communication with \hbox{J.\ Distler and F.\ Zamora}}.'' Unpublished, 1999.

\bibitem{Distler:1999tr}
J.~Distler and F.~Zamora, {\it {Chiral symmetry breaking in the AdS / CFT
  correspondence}},  {\em JHEP} {\bf 05} (2000) 005,
  [\href{http://arxiv.org/abs/hep-th/9911040}{{\tt hep-th/9911040}}].

\bibitem{Pilch:1999misc}
K.~Pilch, ``{\it Notes on perturbative instability of the SO(5),
  SU(2)$\times$U(1)$\times$U(1), and SU(3) AdS$_5$ vacua}.'' Unpublished, 1999.

\end{thebibliography}\endgroup
\bibliographystyle{JHEP}

\end{document}